\documentclass[a4paper,11pt]{article}
\pdfoutput=1 

\usepackage{jheppub-priv}

\hypersetup{
    bookmarksopen,
    pdftitle={Non-Thermal Fixed Point in a Holographic Superfluid},
    pdfauthor={Carlo Ewerz, Thomas Gasenzer, Markus Karl, Andreas Samberg},
    pdfsubject={Holography},
    pdfkeywords={AdS/CFT correspondence, holography, superfluidity, vortices, non-equilibrium}
}

\input{glyphtounicode}
\newcommand{\ie}{i.\,e.}
\newcommand{\eg}{e.\,g.}
\newcommand{\cf}{cf.}

\newcommand{\D}{\text{d}}
\renewcommand{\i}{\text{i}}
\newcommand{\e}{\text{e}}
\newcommand{\LAdS}{L_{\text{AdS}}}
\newcommand{\Zh}{z_{\text{h}}}
\newcommand{\Phit}{\tilde{\Phi}}
\newcommand{\tf}{t_{\text{f}}}
\newcommand{\ti}{t_{\text{i}}}
\newcommand{\lvv}{l_{=}}
\newcommand{\lva}{l_{\ne}}
\newcommand{\Dvv}{D_{=}}
\newcommand{\Dva}{D_{\ne}}
\newcommand{\del}{\partial}

\renewcommand{\Re}{\operatorname{Re}}
\renewcommand{\Im}{\operatorname{Im}}
\renewcommand{\vec}[1]{\boldsymbol{#1}}
\renewcommand{\arg}{\operatorname{arg}}

\newcommand{\eq}[1]{(\ref{eq:#1})}
\newcommand{\Eq}[1]{Eq.~(\ref{eq:#1})}

\newcommand{\Fig}[1]{Fig.~\ref{fig:#1}}
\newcommand{\Tab}[1]{Tab.~\ref{tab:#1}}

\newcommand{\Sect}[1]{Sect.~\ref{sec:#1}}
\newcommand{\Appendix}[1]{App.~\ref{app:#1}}
\newcommand{\Cites}[1]{\cite{#1}}
\newcommand{\Cite}[1]{\cite{#1}}

\title{Non-Thermal Fixed Point in a Holographic Superfluid}

\author[a,b]{Carlo Ewerz,}
\author[a,b,c]{Thomas Gasenzer,}
\author[a,b]{Markus Karl,}
\author[a,b]{Andreas Samberg}
\affiliation[a]{%
  Institut f\"ur Theoretische Physik,
  Ruprecht-Karls-Universit\"at Heidelberg,\\
  Philosophenweg~16,
  69120~Heidelberg, Germany}
\affiliation[b]{%
  ExtreMe Matter Institute EMMI,
  GSI Helmholtzzentrum f\"ur Schwerionenforschung,\\
  Planckstra{\ss}e~1,
  64291~Darmstadt, Germany}
\affiliation[c]{%
  Kirchhoff-Institut f\"ur Physik,
  Ruprecht-Karls-Universit\"at Heidelberg,\\
  Im Neuenheimer Feld 227,
  69120 Heidelberg, Germany}

\abstract{%
  We study the far-from-equilibrium dynamics of a $(2+1)$-dimensional
  super\-fluid at finite temperature and chemical potential using its
  holographic description in terms of a gravitational system in $3+1$
  dimensions. Starting from various initial conditions corresponding to
  ensembles of vortex defects we numerically evolve the system to long
  times.  At intermediate times the system exhibits Kolmogorov scaling
  the emergence of which depends on the choice of initial conditions.
  We further observe a universal late-time regime in which the
  occupation spectrum and different length scales of the superfluid
  exhibit scaling behaviour.  We study these scaling laws in view of
  superfluid turbulence and interpret the universal late-time regime
  as a non-thermal fixed point of the dynamical evolution. In the
  holographic superfluid the non-thermal fixed point can be understood
  as a stationary point of the classical equations of motion of the
  dual gravitational description.
}

\emailAdd{c.ewerz@thphys.uni-heidelberg.de}
\emailAdd{t.gasenzer@uni-heidelberg.de}
\emailAdd{m.karl@thphys.uni-heidelberg.de}
\emailAdd{a.samberg@thphys.uni-heidelberg.de}

\begin{document}

\maketitle
\flushbottom

\section{Introduction}
\label{sec:intro}
Studies of far-from-equilibrium time evolution of quantum many-body systems have intensified considerably during recent years, driven mainly by new technological possibilities.
For example, strongly non-linear dynamics has been observed in ultracold atomic gases \cite{Anderson2001a,Eiermann2004a,Sadler2006a,Weller2008a.PRL101.130401,Weiler2008a,Neely2010a}, or semi-conductor exciton--polariton superfluids \cite{Kasprzak2006a.etal,Lagoudakis2008a,Lagoudakis2009a,Amo2011a}.
Moreover, high-energy heavy-ion collision experiments have brought up
many questions concerning the thermalisation of the quark--gluon
plasma, cf.~\cite{Berges:2012ks} and references cited therein.
Interactions between the constituents of these systems can lead to strong correlations, which render the description of the long-time dynamics intricate and give rise to non-trivial many-body states far from equilibrium.
In the case of strong interactions, a quantitative description of dynamical evolution is typically plagued by the absence of suitable approximation techniques, or by technical difficulties such as sign problems when evaluating the dynamics by means of numerical methods.
Similarly complicated situations can arise even when a weakly
interacting system becomes strongly correlated. In such cases,
non-linear excitations can dominate the system's dynamics such as
solitary waves or topological defects.
In quantum many-body systems the massive appearance of such excitations and their interactions can give rise to quantum turbulence phenomena, \ie, to states the statistical properties of which bear resemblance to correlations observed in classical turbulent systems.

Holographic methods (also known as AdS/CFT or gauge/gravity correspondence) have in recent years
opened new vistas on strongly correlated quantum systems. This approach relies on the
observation that some strongly coupled quantum theories in $D$ dimensions
have a dual (that is equivalent) description in terms of a classical, weakly coupled gravitational
theory in $D+1$ dimensions. Loosely speaking, the quantum theory can be thought
of as living on the $D$-dimensional boundary of the higher-dimensional space.
One therefore often speaks of the `boundary theory' when addressing the
$D$-dimensional quantum system described in the holographic framework.
A finite temperature of the quantum system
has a holographic description on the gravity side in terms of a black hole with
the corresponding Hawking temperature. The original discovery of the holographic duality
\cite{Maldacena:1997re,Gubser:1998bc,Witten:1998qj} was made for the case of a 4-dimensional
gauge theory with maximal supersymmetry. It has subsequently been realised that the method
is far more general and can be applied to a variety of physical systems. Applications of the
holographic duality by now comprise topics as diverse as ultracold quantum gases and
the quark--gluon plasma, for a review see for example \cite{Adams:2012th}.
A novel and exciting development is the extension of the holographic duality to
various condensed-matter quantum systems, for reviews see for instance
\cite{Hartnoll:2009sz,McGreevy:2009xe}.
The holographic duality allows one to study the dynamical real-time evolution
of a strongly correlated many-body quantum system in a genuinely nonperturbative
framework. Remarkably, the intricate dynamics of the system -- including
its far-from-equilibrium behaviour -- is entirely captured by a classical
gravitational system. Evidently, the holographic description therefore
offers the potential to address phenomena in the quantum system which are
notoriously difficult to access by other methods. It even carries the promise
to discover new phenomena that are unique to strong-coupling situations.

A particularly interesting discovery is that there are gravitational systems which
have a dual interpretation in terms of superconductors or superfluids in $2+1$
dimensions \Cites{Gubser:2008px,Hartnoll:2008vx,Herzog:2008he}.
Here, the dual gravitational description is in terms of an Abelian Higgs model
on and coupled to a $(3+1)$-dimensional spacetime of negative cosmological constant, an Anti-de Sitter spacetime. The breaking of the Abelian $U(1)$ symmetry
at low temperature is associated with the condensation of an order-parameter field
and can be interpreted as the emergence of superconductivity or superfluidity.

In the present paper, we will consider a holographic superfluid of this kind at finite temperature
and with a chemical potential for the $U(1)$ charge. In this system, vortex excitations exist
in the superfluid phase without the presence of an external magnetic field.\footnote{This is
opposed to the case of a holographic superconductor. For a more detailed discussion
of the differences between holographic superfluids and holographic superconductors
in view of the corresponding vortex solutions see for example \cite{Dias:2013bwa}.}
The system we consider is a $(2+1)$-dimensional relativistic superfluid.
Studies of various aspects of that particular holographic superfluid include
\cite{Keranen:2009re,Bhaseen:2012gg,Dias:2013bwa}.
A study of its time-evolution as it relaxes from a
far-from-equilibrium initial state, corresponding to an ensemble of
vortex defects, was performed in \cite{Adams:2012pj}. There, the
authors numerically solve the gravitational Einstein--Maxwell--scalar
system dual to this superfluid.
They have identified a certain
regime in the evolution in which the superfluid exhibits Kolmogorov scaling.
In the present paper, we perform a similar numerical analysis of the same system.
As in \cite{Adams:2012pj} we treat the holographic superfluid in the
so-called probe limit in which the AdS spacetime in the dual gravitational description
is kept fixed. Our numerical methods are sufficiently fast to allow us to investigate
two new aspects of the far-from-equilibrium evolution of the superfluid.
On the one hand, we follow the system's evolution for a very long time.
On the other hand, we study various initial conditions, in particular, we choose random
distributions as well as lattices of vortex defects of different densities. This makes
it possible to clearly identify the time scales at which the system enters a
universal regime.

In particular the investigation of the late-time behaviour of the system leads us
to a very interesting observation. Following the propagation and annihilation
of the quantum vortices in time, we are able to observe a stage of universal
critical dynamics arising in the late-time evolution of the superfluid when
the quantum turbulent ensemble relaxes towards equilibrium.
More specifically,
we observe how the system approaches a non-thermal fixed point, \ie, a far-from-equilibrium field configuration exhibiting universal scaling behaviour \cite{Berges:2008wm,Berges:2008sr}.
We demonstrate how the approach to this fixed point can be related to the dynamics of vortex defects in the order parameter.
Non-thermal fixed points were identified, in quantum field theory, as stationary solutions of non-perturbative equations of motion for Green functions \cite{Berges:2008wm,Scheppach:2009wu,Berges:2010ez}.
In the context of non-relativistic Bose gases as well as of relativistic scalar and gauge theories it was shown that superfluid turbulence, related to characteristic distributions of vortex defects \cite{Nowak:2010tm,Nowak:2011sk} or more general non-linear excitations  \cite{Karl:2013mn,Gasenzer:2011by,Gasenzer:2013era}, can be interpreted in terms of non-thermal fixed points.
General universality classes of such non-thermal fixed points are expected to emerge from a renormalisation-group analysis which includes scaling in space and time \cite{Gasenzer:2008zz,Mathey2014a.arXiv1405.7652}.

While non-thermal fixed points have been discussed in various contexts
their properties in strongly coupled systems have not yet been
explored.
The present study is the first analysis of the holographic superfluid in view of the approach to a non-thermal fixed point, comparing universal and non-universal stages of the superfluid's evolution.
It opens a new and exciting perspective on time evolution as described in a holographic setting, in particular on the mutual implications of such universal dynamics on both the gravity and the boundary-theory sides.

Our paper is organised as follows.
In \Sect{holo-superfluid} we summarise the definition and the properties of the Einstein--Maxwell--scalar gravity model dual to the superfluid, as well as the implementation of the resulting  equations within our numerical approach.
\Sect{non-eq-dynamics} contains the details about the different initial conditions considered and summarises basic properties of vortices in superfluids.
Further in that section, we present our numerical results on the evolution of the vortex characteristics and of the statistical properties of the ensembles, in particular on occupation number spectra of the boundary theory.
In \Sect{hNTFP} we finally discuss the holographic perspective on the non-thermal fixed points in the superfluid's
evolution.
A summary and outlook in \Sect{summary} complete the main presentation. Some technical
considerations are given in two appendices. Appendix \ref{app:NumImpl} contains
details concerning the numerical implementation, while Appendix \ref{app:determ-scal-law}
deals with the method for extracting scaling exponents from various spectra.

%

\section{Holographic superfluid}
\label{sec:holo-superfluid}
In this section, we set out the holographic framework for
describing the dynamics of a superfluid in $(2+1)$ dimensions by means
of a gravitational dual in a $(3+1)$-dimensional spacetime.

\subsection{Gravity model dual to the superfluid}
\label{sec:model}
The holographic framework for superfluidity was laid down in \Cites{Gubser:2008px,Hartnoll:2008vx,Herzog:2008he}.
A scalar field is dynamically coupled to an Abelian gauge theory and gravity on a $(3+1)$-dimensional spacetime. We use the action
\begin{equation}
  \label{eq:47}
  \begin{gathered}
    S = \frac{1}{2\kappa} \int \D^4 x \sqrt{-g}\left(\mathcal{R} - 2\Lambda + \frac{1}{q^2}\mathcal{L}_{\text{matter}}\right) \,,\\
    \mathcal{L}_{\text{matter}}= -\frac{1}{4} F_{MN}F^{MN} - \left\lvert D_M\Phi\right\rvert^2 - m^2 \lvert\Phi\rvert^2 \,.
  \end{gathered}
\end{equation}
Here, $\kappa$ is the Newton constant in four dimensions, and the
cosmological constant $\Lambda=-3/\LAdS^2$ is negative. $\LAdS$
sets the curvature scale of the spacetime which arises as a solution of the
corresponding Einstein equations.
$\mathcal{R}$ is the Ricci scalar constructed from the
metric $g_{MN}$, and $g$ is the determinant of that metric.\footnote{We
  use capital Latin indices for coordinates of the 4-dimensional bulk
  spacetime, $M,N=t,x,y,z$, and Greek indices for the 3-dimensional
  spacetime of the dual boundary theory, $\mu,\nu=t,x,y$.}
The Lagrangian density $\mathcal{L}_{\text{matter}}$ accounts for the gauge fields,
with the associated field-strength tensor $F_{MN}=\nabla_M A_N - \nabla_N A_M$, and for the scalar field $\Phi$.
Here, $\nabla_M$ denotes the covariant derivative associated with the Levi-Civita
connection.
The local $U(1)$ gauge symmetry of the Lagrangian is implemented by upgrading $\nabla_M$ to the gauge-covariant derivative $D_M = \nabla_M-\i A_M$.

According to the holographic dictionary the gauge potential  $A_M$ in the bulk induces a global $U(1)$ symmetry of the dual field theory.
The operator dual to $A_M$ is the conserved $U(1)$ current $j^\mu$ which arises from that symmetry, see \Appendix{EOMs} for details.
Finally, the complex scalar field $\Phi$ has mass $m$ and charge $q$ which, by suitable rescaling of the fields, has been pulled out of $\mathcal{L}_{\text{matter}}$.
If the gravity is taken to be dynamic, $q$ thus quantifies the coupling between the gravity and gauge--matter parts of the model.

Solving the gravity model \eq{47} allows us to obtain information about the dynamical evolution of a superfluid described by the boundary theory.
A superfluid is commonly described by a complex scalar field $\psi$ which, in the symmetry-broken phase, assumes a non-vanishing expectation value $\langle\psi\rangle \ne 0$ reflecting the presence of a Bose--Einstein condensate.
In our holographic model, the field operator $\psi$ is dual to the scalar field $\Phi$, as we explain in detail in \Appendix{EOMs}.
The solutions of the equations of motion of the model \eqref{eq:47} are subject to boundary conditions in the holographic direction.
These conditions determine the temperature and the chemical potential for the $U(1)$ charge $j^{0}$ in the $(2+1)$-dimensional boundary theory.
In particular, the temperature and chemical potential can be chosen such that the boundary theory is in the symmetry-broken phase with a condensate, $\langle\psi\rangle\ne 0$ \cite{Gubser:2008px}.
Holography allows us to compute the time evolution of the quantum expectation value $\langle\psi\rangle$ starting from various far-from-equilibrium states by solving the classical dynamics of \eqref{eq:47}.
In fact, we also have access to the phase angle of the complex field $\langle\psi\rangle$ the spatial variation of which encodes information about the superfluid flow.
We will use this to construct far-from-equilibrium initial states.

In this article, we consider the so-called probe limit of the action \eqref{eq:47} in which the back-reaction of the fields $\Phi$ and $A_M$ on the metric is neglected.
This is a good approximation for large scalar charge $q$, as is clear from the rescaled form of the action \eqref{eq:47}, with $1/q^{2}$ entering as a small pre-factor of the gauge--matter Lagrangian, see for example \cite{Hartnoll:2008vx,Albash:2009iq}.
We can thus treat the gravity and matter parts separately from each other.

Ignoring for the moment the gauge--matter part of the model, the Einstein equations are solved
by an AdS$_4$ spacetime with a planar Schwarzschild black hole.
The respective metric reads
\begin{equation}
  \label{eq:metric}
  \D s^2 = \frac{\LAdS^2}{z^2} \left( -h(z)\D t^2 + \D\vec{x}^2 - 2\D t\,\D z \right) \,,
\end{equation}
with the horizon function
\begin{equation}
  \label{eq:6}
  h(z) = 1 - \left(\frac{z}{\Zh}\right)^3 \,.
\end{equation}
Here, $(t,\vec{x})=(t,x,y)$ are the coordinates of the spacetime on
which the {boundary} field theory is defined, and $z\geq0$ is the
coordinate of the holographic direction.
As already pointed out, it is often useful to think of the dual field theory dynamics to take place at the boundary  $z=0$.
However, it is the entire bulk information that encodes the boundary physics.
The black-hole horizon is situated at $z=\Zh$.
We use Eddington--Finkelstein coordinates, where a light ray falling into the black hole is affinely parameterised by $z$, with all other coordinates kept constant.
Such coordinates are particularly well-suited for the numerical solution of the system's real-time dynamics \cite{Adams:2012pj} and are often employed in holography \cite{Chesler:2013lia}.

Working, then, with a fixed background metric, one is left with the equations of motion for the matter part, \ie, the generally covariant Maxwell and Klein--Gordon equations which are coupled to each other through the bulk electromagnetic current.
As the background metric, together with the gauge coupling, allows for spontaneous symmetry breaking in the scalar sector, the problem has thus been reduced to solving a classical Abelian Higgs
model on the curved background \eqref{eq:metric}.

The holographic model captures important aspects of Tisza's
two-fluid model \cite{Tisza1938TPiHII}, for a review see for instance
\cite{tilley1990superfluidity}.
More specifically, it has been shown that in
the hydrodynamic expansion, to an order which includes only non-dissipative terms, the
boundary theory reduces to a relativistic
version of the two-fluid model \cite{Sonner:2010yx}.
We point out that our treatment of the holographic model captures more
than the effective hydrodynamic limit of the boundary theory. In fact, the holographic
description is valid at all scales. Nevertheless, it can still be
useful to think of the dynamics of the superfluid in terms of two
distinct components.
In the probe limit, the presence of the static black hole at $z=\Zh$
translates, by the AdS/CFT dictionary, to a static heat bath of
temperature $T = 3/(4\pi\Zh)$ in the boundary theory.  Loosely speaking, this can be
viewed as the normal component of the fluid.
Similarly, the fields $\Phi$ and $A_M$ holographically represent the superfluid component.
The superfluid component is coupled to the normal component and can dissipate energy and momentum to it.
Thus, the model naturally incorporates dissipation to a thermal bath.
Further details of the interpretation of the probe limit can be found in \Cite{Adams:2012pj}.
We remark that our formalism is manifestly relativistic.
The superfluidity described by the boundary field theory appears in the non-relativistic  low-energy limit \cite{Adams:2012pj}.

\subsection{Implementation}
\label{sec:implementation}

In this article, we consider the time evolution of the superfluid starting from a far-from-equilibrium situation.
For this, we need to numerically solve the full equations of motion for the gauge field $A_M$ and the scalar field $\Phi$ in the bulk,
\begin{gather}
  \nabla_M F^{MN} - \i \left(\Phi^* D^N \Phi - \Phi \left[D^N \Phi\right]^* \right) = 0 \,,\label{eq:1}\\
  \left(D^2 - m^2 \right)\Phi = 0\,.\label{eq:8}
\end{gather}
By the AdS/CFT dictionary, we can extract the expectation value $\langle\psi(\vec{x}, t)\rangle$, \ie~the superfluid order parameter, from the asymptotic behaviour of the dual scalar field $\Phi$ close to the boundary at $z=0$, see \Appendix{EOMs}.

We choose the temperature and chemical potential in the boundary field theory such that the system is in the superfluid phase.
Our units are fixed by setting $\Zh=1$ in the metric \eqref{eq:metric}, such that the temperature is $T= 3/(4\pi)$.
Then, choosing a chemical potential of $\mu=8\pi T$ puts the system into the superfluid phase, at a temperature $T/T_{\text{c}} = 0.68$.

As we aim at studying universal aspects of the superfluid, we consider different types of initial conditions, containing topological defects, the details of which are discussed in  \Sect{initialCond}.
We take the $(x,y)$ directions to be periodic and use a pseudo-spectral basis for the fields.
In order to be able to properly do statistics and suppress finite-volume effects we need to choose our numerical grid sufficiently large.
Specifically, we study grid sizes of $352\times 352$ as well as $504\times 504$ points in the $(x,y)$ plane.
The data from these different grid sizes are qualitatively consistent.
Since observables computed on the larger grids are considerably less noisy, all data presented in the following were produced on $504\times 504$ grids.
For this choice, the dimensionless product $LT$ of the extent of the $(x,y)$-domain and the temperature is $LT = 34.4$.
We use a basis of 32 Chebyshev polynomials and 32 grid points in the holographic direction.
To be able to properly assess the genuine late-time behaviour of the system we let it evolve to time $\tf = 4000$ in the aforementioned units, or $\tf T = 955$ in units of temperature.
We use an explicit time-stepping scheme for the propagation.
We point out that the timestep $\tau$ used in our numerics is much smaller than one unit of time, $\tau \ll 1$.
A more detailed discussion of our choice of numerical methods and parameters is given in \Appendix{NumImpl}.

For technical reasons, each of our initial conditions depends on some random data.
To get more robust results that do not depend on the definite values of these random data in a specific realisation, we average statistical observables over ten runs for each type of initial condition (random distributions and lattices of vortices, see \Sect{initialCond}).

\section{Holographic non-equilibrium dynamics}
\label{sec:non-eq-dynamics}

In this section, we first discuss the different types of initial
conditions that we employ to induce non-equilibrium behaviour of the
superfluid.
Specifically, our initial conditions are characterised by ensembles of topological vortex defects.
Then, we analyse the evolution of the system in terms of the distribution statistics of the defects.
Furthermore, we consider the occupation spectra of the superfluid order parameter.

\subsection{Initial conditions}
\label{sec:initialCond}

For the quantum systems we have in mind, quenches, especially across a
or in the vicinity of a phase transition, are being studied
intensively, both experimentally and theoretically.  A quench
in the usual sense involves the rapid change of either a thermodynamic
parameter, such as temperature, or a Hamiltonian parameter, for
example interaction strength.
Temperature quenches across a superfluid phase transition were
recently studied in the holographic approach, in the context of the
Kibble--Zurek scenario \cite{Sonner:2014tca,Chesler:2014gya}.
Novel techniques developed for ultracold quantum gases and
exciton--polariton superfluids allow to rapidly change parameters of the Hamiltonian.
Using these, ensembles of vortex defects can be created in the superfluid
\cite{Anderson2001a,Eiermann2004a,Sadler2006a,Weller2008a.PRL101.130401,%
Weiler2008a,Neely2010a,Kasprzak2006a.etal,Lagoudakis2008a,Lagoudakis2009a,Amo2011a}.
The generic consequence of both types of quenches for
superfluids in two spatial dimensions is the nucleation of quantised
vortices.
This behaviour was also observed in simulations of both relativistic
\cite{Gasenzer:2011by,Gasenzer:2013era} and non-relativistic
\cite{Nowak:2010tm,Nowak:2011sk,Schole:2012kt,Karl:2013mn,Karl:2013kua}
Bose systems.
Here, we directly prepare ensembles of vortices as quench-like initial
conditions for the superfluid's time evolution.
For simplicity, we refer to our initial conditions as quenches in the following.

The structure of the core of a vortex, as well as collective properties of the ensemble,
are specific characteristics of the superfluid system.
The local phase structure of the superfluid order parameter around a vortex defect, however,
is determined by topology. The presence of a local vortex
with quantisation $w\in\mathbb{Z}\backslash \{0\}$ requires that the
phase angle $\varphi=\operatorname{arg}(\langle\psi\rangle)$ of the
superfluid order parameter has winding number $w$
around the vortex, \ie, $\oint \D\varphi = 2\pi w$, where
the integration contour encircles the particular defect.  This property can be
used to prepare the initial-time order-parameter field to bear a set
of vortices ($w>0$) and anti-vortices ($w<0$).  We do this by
`multiplying' a localised vortex into the phase of the thermalised
superfluid, $\langle\psi\rangle \to \langle\psi\rangle \cdot \e^{\i w
  \phi_{\text{v}}}$ with an appropriate polar angle $\phi_{\text{v}}$.
During the initial propagation of the equations of motion, the
appropriate density profile $|\langle\psi\rangle|^{2}$ around the
defect builds up in a short time, providing us with an ensemble of
vortices on top of a previously equilibrated system. Technical details
can be found in \Appendix{EOMs}.  Previous studies of
vortex solutions in holographic superfluids include
\cite{Keranen:2009re,Dias:2013bwa}.

\begin{figure}[!t]
  \centering
  \includegraphics[width=0.7\textwidth]{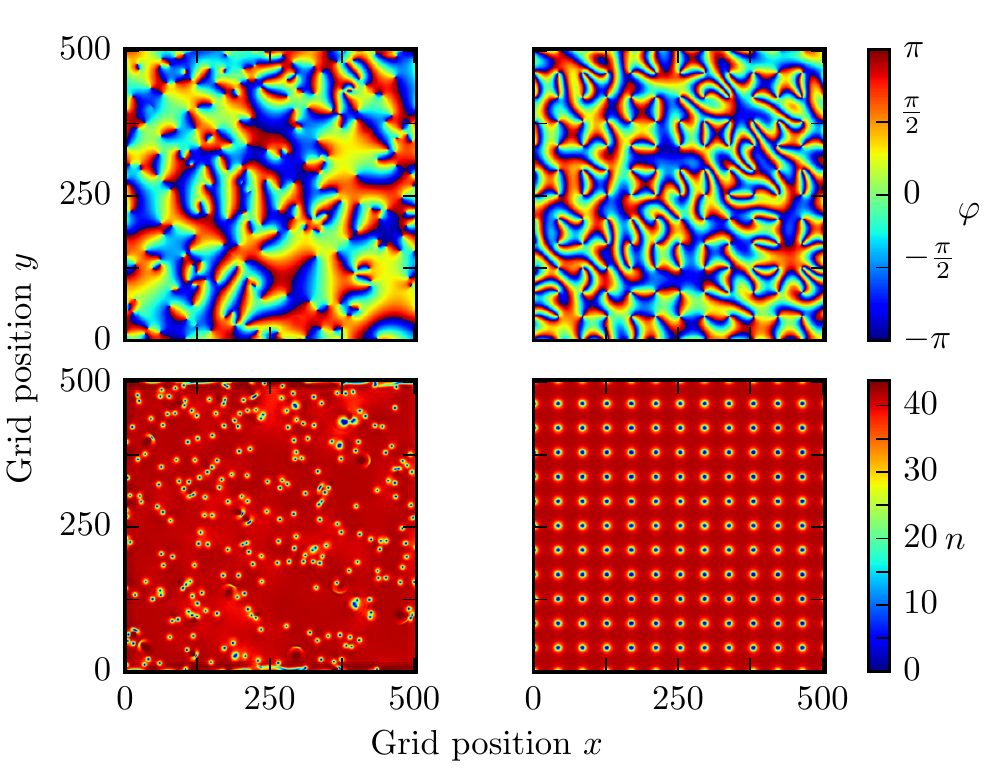}
  \caption{%
    Illustration of two classes of initial conditions we have chosen
    in this work. %
    Left column: random distribution with an equal number of
    elementary vortices of either sign. 
    Right column: vortex lattice with winding numbers $\pm 2$
    alternating from site to site. 
    Shown are single-run snapshots of the phase-angle distributions
    $\varphi(\vec{x}) = \arg(\langle\psi(\vec{x})\rangle)$ (first row)
    and densities
    $n(\vec{x})=\lvert\langle\psi(\vec{x})\rangle\rvert^2$ (second
    row) of the superfluid order-parameter field
    $\langle\psi(\vec{x})\rangle$ at time $t=0$ when the vortex cores
    are fully developed. %
    The noise that is at this time added onto the phase distribution
    for the vortex lattices is clearly visible.%
    \label{fig:initial-conditions}
  }
\end{figure}
%
In \Fig{initial-conditions} we show example realisations of different
types of initial conditions we prepare, with a random distribution
of vortices and anti-vortices (left column) and a regular lattice
distribution (right column).
The graphs in the first row show the phase angle field
$\varphi(\vec{x})$.
After imprinting the phase winding of the vortices the system
reacts by building up vortex cores and, on a longer time scale,
by redistributing the defects, as a consequence of
interactions in the superfluid. Thus, the short-time outcome of the respective
quench is defined by choosing number, quantisation, and spatial distribution of
the vortices.
The second row in \Fig{initial-conditions} shows that indeed  vortex cores
are built up shortly after the winding phases have been imprinted.
This procedure corresponds to the well-established technique of
imprinting vortices by `stirring' a Bose--Einstein
condensed dilute atomic gas with the help of a laser
\cite{Raman2001a.PRL.87.210402R,Abo-Shaeer2001a}.
Experimental methods have been refined for creating
vortices in cold atomic gases \cite{Weiler2008a,Neely2010a,Neely2012a,Kwon2014a.arXiv1403.4658K}
and exciton--polariton superfluids \cite{Lagoudakis2008a,Lagoudakis2009a}.

In this work, we study two classes of initial vortex distributions:
class $\mathcal{A}$ consists of random distributions of elementary
vortices of winding numbers $\pm 1$, while class $\mathcal{B}$
comprises regular $12\times 12$ lattices of non-elementary vortices
with absolute winding numbers $\lvert w\rvert>1$, alternating in sign
from site to site.  For examples from each class see
\Fig{initial-conditions}.
We observe that a non-elementary vortex of absolute
winding number $w$ quickly decays into $w$ elementary vortices of
the same sign.
Therefore, there are strong correlations built into the initial vortex
positions for initial conditions in class $\mathcal{B}$, insofar as
after the decay of the non-elementary vortices the like-sign singly
quantised vortices are clustered. On the other hand, within the
limitations of the finite grid, the initial vortex positions in class
$\mathcal{A}$ are completely uncorrelated.
For each class of initial conditions, we vary the number of
vortices by randomly distributing $144, 432$, or $720$ vortices of either
sign in class $\mathcal{A}$, and choosing winding numbers $\pm 2,
\pm 6$, or $\pm 10$ in class $\mathcal{B}$.
This choice implies that after the decay of the
non-elementary vortices the total number of elementary vortices is the
same in the corresponding cases of both classes.
In this way, we vary the initial vortex correlations and the mean separation of vortices,
considering both as quench parameters.
The resulting six different initial conditions are summarised
in \Tab{init-conds}. Note that in all cases the net
vorticity is zero.
%
\begin{table}[t]
  \centering
  \begin{tabular}{l|ccc}
    & I             & II            & III          \\
    \hline
    $\mathcal{A}$\hspace*{1ex} random distribution & $2\times 144$ & $2\times 432$ & $2\times 720$\\
    $\mathcal{B}$\hspace*{1ex} vortex lattice ($12\times 12$) & $\pm 2$ & $\pm 6$ & $\pm 10$\\
    \hline
    \# elementary vortices & 288 & 864 & 1440
  \end{tabular}
  \caption{%
    Vortex numbers and winding numbers for the six different types of initial distributions used in the
    simulations. Class $\mathcal{A}$ consists of random distributions of three different numbers of
    elementary vortices and anti-vortices, while class $\mathcal{B}$ comprises regular square lattices
    of alternating-sign non-elementary vortices with three different winding numbers.
    See \Fig{initial-conditions} for example realisations of types $\mathcal{A}$\,I and $\mathcal{B}$\,I.%
    \label{tab:init-conds}
  }
\end{table}

The vortex phases are imprinted at a time $t=\ti<0$. After starting
the simulation, stable vortex cores develop quickly, typically after
$\Delta t = 5$ and $\Delta t = 10$ for
quenches of class $\mathcal{A}$ and $\mathcal{B}$, respectively. We
adjust $\ti$ such that at $t=0$ the vortex cores are fully
formed.
For class $\mathcal{B}$ (vortex lattices), we furthermore perturb the
phase of the bulk scalar field at time $t=0$ with random noise to
induce variations in the decay pattern. This is illustrated in the
upper right panel in \Fig{initial-conditions}. For class $\mathcal{A}$
(random distribution) it is not necessary to add phase noise due to
the randomness of the vortex positions.
%

\subsection{General considerations on vortex dynamics in a superfluid}
\label{sec:gener-cons-vort}

It will be useful to discuss the real-time dynamics of our holographic superfluid
in the framework of an effective description that is well known from other superfluid
systems. In this effective picture of quantum turbulence the vortices appear as
collective excitations of the order-parameter field. In addition to the vortices the
system contains sound waves. These also mediate the effective interaction of the vortex defects.
The sound waves can usually be treated in a good approximation as linear perturbations of
the order-parameter field.

In this subsection we summarise a few basic properties of the effective description known to characterise vortices and their dynamics in two-dimensional non-relativistic superfluids.
For reviews in the context of superfluid helium and cold atomic gases see, \eg, \cite{Donnelly1991a,Inguscio1999a,Tsubota:vortices}.
The time evolution of an undamped dilute non-relativistic superfluid carrying vortex defects
is understood to be well described
by the non-linear Schr\"odinger or Gross--Pitaevskii equation (GPE)
\cite{Gross1961a,Pitaevskii1961a}, the classical field equation for the order parameter $\langle\psi\rangle$ with Schr\"odinger-type free and quartic interaction parts.
The potential field $\vec{v}\sim\vec{\nabla}\varphi$ derives from the order parameter's phase angle $\varphi=\arg(\langle\psi\rangle)$ and
describes the local velocity of the superfluid.
The velocity field is thus curl free.
Vorticity is carried rather by the vortex defects at which the
order-parameter field vanishes, permitting a finite circulation of the phase
around it.

In studies of two-dimensional turbulence, important observables are
the kinetic energy spectrum $\propto\langle[\vec{v}(\vec{k})]^{2}\rangle$
and the distribution of vorticity $\omega=\partial_xv_y-\partial_yv_x$
which derives from the local velocity field $\vec{v}=(v_x,v_y)$.
As was first discussed in the context of classical fluid dynamics
\cite{Lagally1921a,Lin1941a,Onsager1949a},
it is convenient to think of vortices as Coulomb-interacting
`charges' in an effective `electrostatic' picture where vortices of
opposite-sign (like-sign) winding number attract (repel) each
other.
The dynamics of the GPE vortices is to a good approximation of
Hamiltonian character, where the position along one spatial dimension
forms the canonical momentum of the position along the other.
This forces, in effect, oppositely charged vortices to move in parallel at a fixed distance
(Helmholtz law) while equal-sign vortices circulate around each other.
The statistics of the vortex distribution contains important
information about the temporal and spatial characteristics of the system.

Bogoliubov sound waves form the weak linear excitations of the order parameter field,
through which vortices can interact with each other.
The interactions with a fluctuating, \eg~thermal, background of
excitations of the superfluid causes the vortices to show deviations from
the Hamiltonian behaviour.
For instance, a sufficiently strong dissipative force can suppress the Helmholtz pair propagation
and make oppositely charged vortices move towards each other.
This is well known for defect solutions of Ginzburg--Landau equations which represent
the generalisation of the GPE with complex parameters \cite{aranson2002a}
and are of relevance, beyond superfluids, for many applications including liquid
crystals and biological systems.

On a mean-field level, aspects of vortex annihilation in a superfluid
can also be described in terms of phase-ordering kinetics
\cite{bray1994}. In this context, it is assumed that in the
`coarsening regime' the system can be characterised by a single length
scale exhibiting a scaling law with respect to time.
However, cases have been identified \cite{Damle1996,Nazarenko2006} in
which systems can deviate from the simple scaling predictions of
\cite{bray1994}.
In the present paper, we study non-thermal fixed points
\cite{Berges:2008wm} of which phase-ordering kinetics represents one
possible realisation.  Note, however, that the concept of non-thermal
fixed points reaches beyond phase-ordering kinetics.  For example,
they can be associated with turbulent processes which can have effects
opposite to ordering kinetics.  In the language of turbulence,
ordering kinetics corresponds to an inverse cascade directed from
small to large length scales, building up large-scale correlations in
the system.  In contrast, a direct cascade transports energy in the
opposite direction, creating small-scale fluctuations.  In
\Sect{hNTFP}, we discuss non-thermal fixed points in the context of
our findings.  In fact, it is possible that the dynamics we discuss in
this work contains both, ordering processes, \ie\ inverse cascades,
and direct turbulent cascades directed from large to small length
scales.

We emphasise that not all of the typical properties mentioned above
will necessarily be seen in the holographic superfluid that we study
here.
But we find it helpful to compare our findings to those properties in
the following.

\subsection{Dynamical evolution of the vortex ensembles in the holographic superfluid}
\label{sec:stag-dynam-evol}

We now turn back to our holographic superfluid. In the following we discuss
the dynamics of the holographic superfluid for $t>0$,
after imprinting the initial conditions discussed in \Sect{initialCond}.
The vortices start to move around, subject to an effective interaction.
When vortices of opposite sign approach each other sufficiently
closely they mutually annihilate.
There appears to be a strong suppression of Helmholtz pair
propagation: The vortex' and the anti-vortex' trajectories only bend
slightly in a common direction before annihilating.
This is an indication of strong dissipative effects in the holographic superfluid.
Annihilations reduce the total vortex number and change
length scales of the vortex distribution. This has
important implications for the turbulence properties of the system,
which we discuss in \Sect{critical-dynamics}.
At the time $\tf=4000$ where we end our simulation, only very few
vortices are left, typically about 4 to 8 for all types of quenches
considered.  In our simulations, the non-zero temperature of the black
hole is related to dissipation such that sound fluctuations are quickly
damped out.
The spontaneous generation of vortex--anti-vortex pairs could not be observed.
We expect that, after all vortices have annihilated, the excess energy
which we initially injected into the superfluid component is completely
dissipated into the heat bath.
As a consequence, the condensate $\langle\psi(\vec{x},t)\rangle$ would relax to a
homogeneous, fully ordered state.
Within the finite evolution times, our simulations give indications for this behaviour.

%
\begin{figure}[!t]
  \centering
  \includegraphics[width=0.7\textwidth]{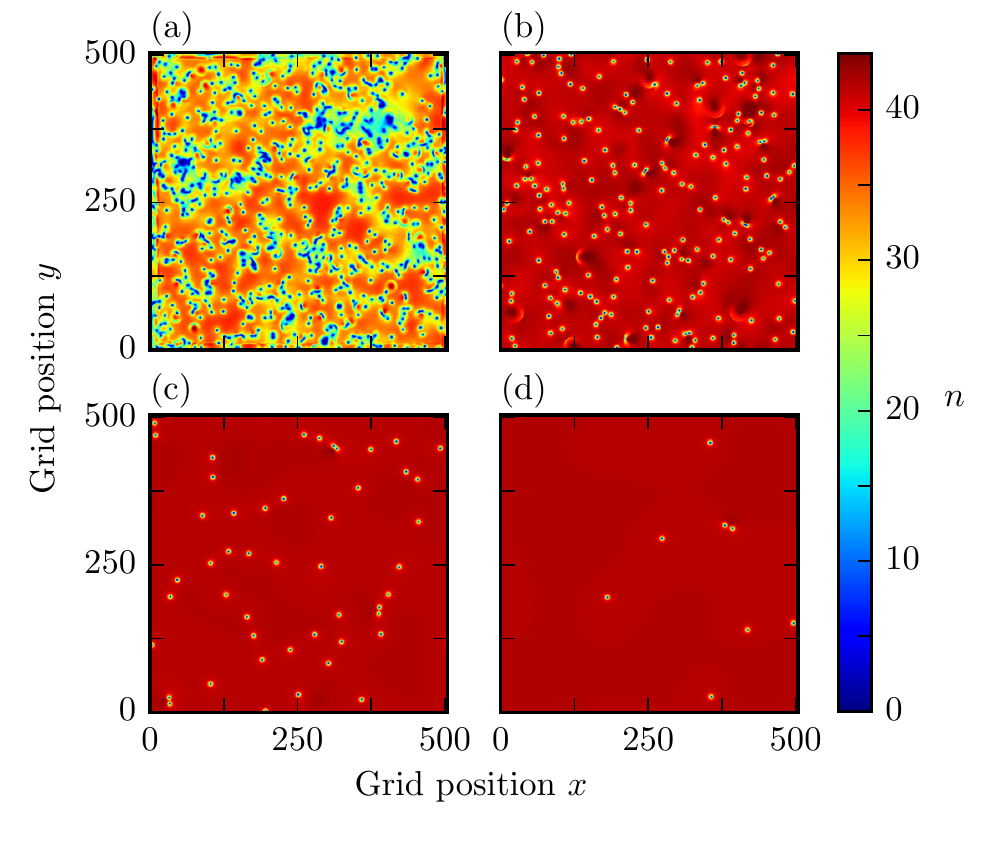}
  \includegraphics[width=0.7\textwidth]{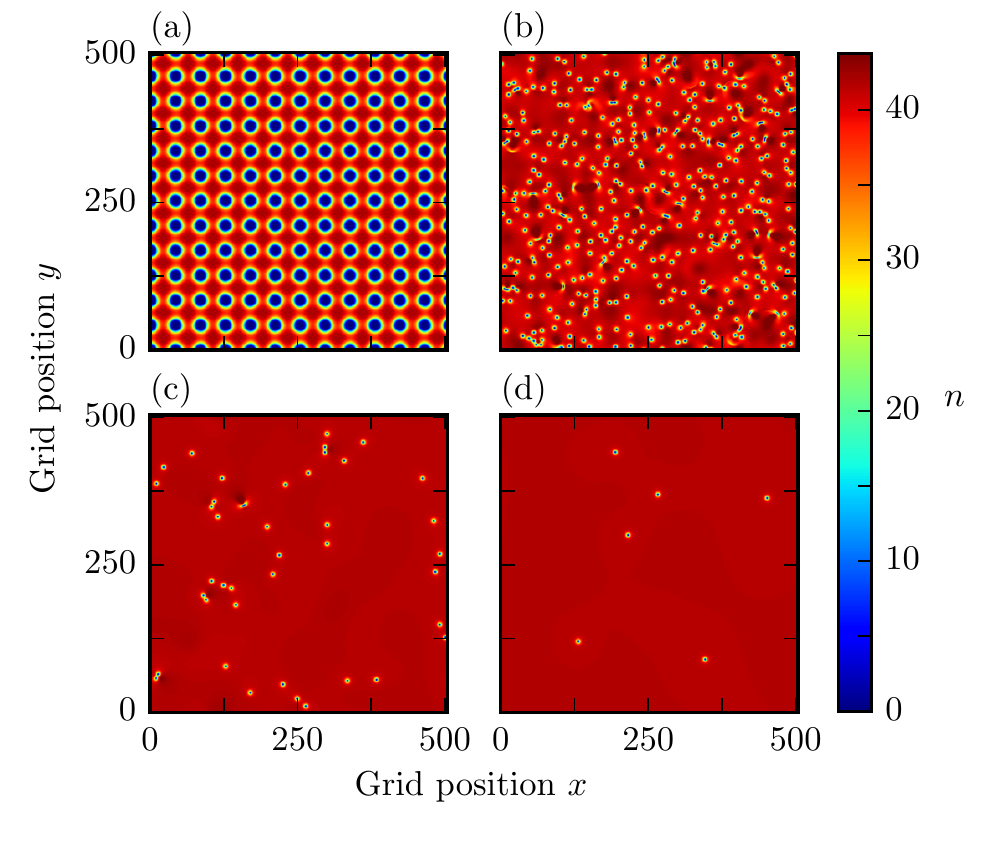}
  \caption{%
    Single-run snapshots of the superfluid density $n(\vec{x}) =
    \lvert\langle\psi(\vec{x})\rangle\rvert^2$ showing the characteristic
    stages of the evolution of the system, after a quench of type
    $\mathcal{A}$\,II (random distribution with 432 elementary
    vortices of either sign, upper panel) and after a quench of type
    $\mathcal{B}$\,II (vortex lattice with winding numbers $\pm 6$,
    lower panel). %
    The vortices can be discerned as dips in the density, and one can
    observe sound waves from vortex annihilation events. The
    snapshots are taken at times (a) $t=0$, (b) $t=100$, (c) $t=600$,
    and (d) $t=4000$. The quenches have been performed at times $t =
    -5$ (quench $\mathcal{A}$\,II) and $t=-10$ (quench
    $\mathcal{B}$\,II).%
    \label{fig:dens-evol}
  }
\end{figure}
%
Snapshots of the full time evolution of the system for the two classes of initial
conditions are shown in \Fig{dens-evol}, where the upper (lower) panels
correspond to a random, \ie, class-$\mathcal{A}$ (class-$\mathcal{B}$, lattice) initial distribution.
We plot the superfluid density $n(\vec{x}) = \lvert\langle\psi(\vec{x})\rangle\rvert^2$ at various times. These
snapshots are representative for the different stages encountered in the
evolution.%
\footnote{\url{http://www.thphys.uni-heidelberg.de/holographic-superfluid} links to videos of example evolutions.\label{fn:movies}}
In both cases, the main features of the dynamic evolution can be
understood as due to the motion and annihilation of vortices. Due to the relatively
high initial density of vortices the first stage of the evolution is
characterised by a small mean vortex--anti-vortex distance and a high annihilation
rate.
Each annihilation event releases energy in form of sound waves
which is then dissipated into the thermal background \cite{Adams:2012pj}.
For initial configurations of class $\mathcal{B}$, there is an additional stage in which
non-elementary vortices decay.
The emerging elementary vortices drift apart for a certain amount of time, thereby
expanding like-sign clusters.
Naturally, these early stages are highly parameter-dependent and therefore non-universal.
The following stage is an evolving dilute vortex gas.
Although it is still the annihilation of vortex--anti-vortex pairs which
brings the system closer to equilibrium, essential aspects of the time
evolution of the system at this stage can be understood from the
statistics of the vortex distribution as we will discuss below.
Eventually, when only one pair remains, the evolution is governed by
the interactions with the sound modes.

\begin{figure}[!t]
  \centering
  \includegraphics[width=0.62\textwidth]{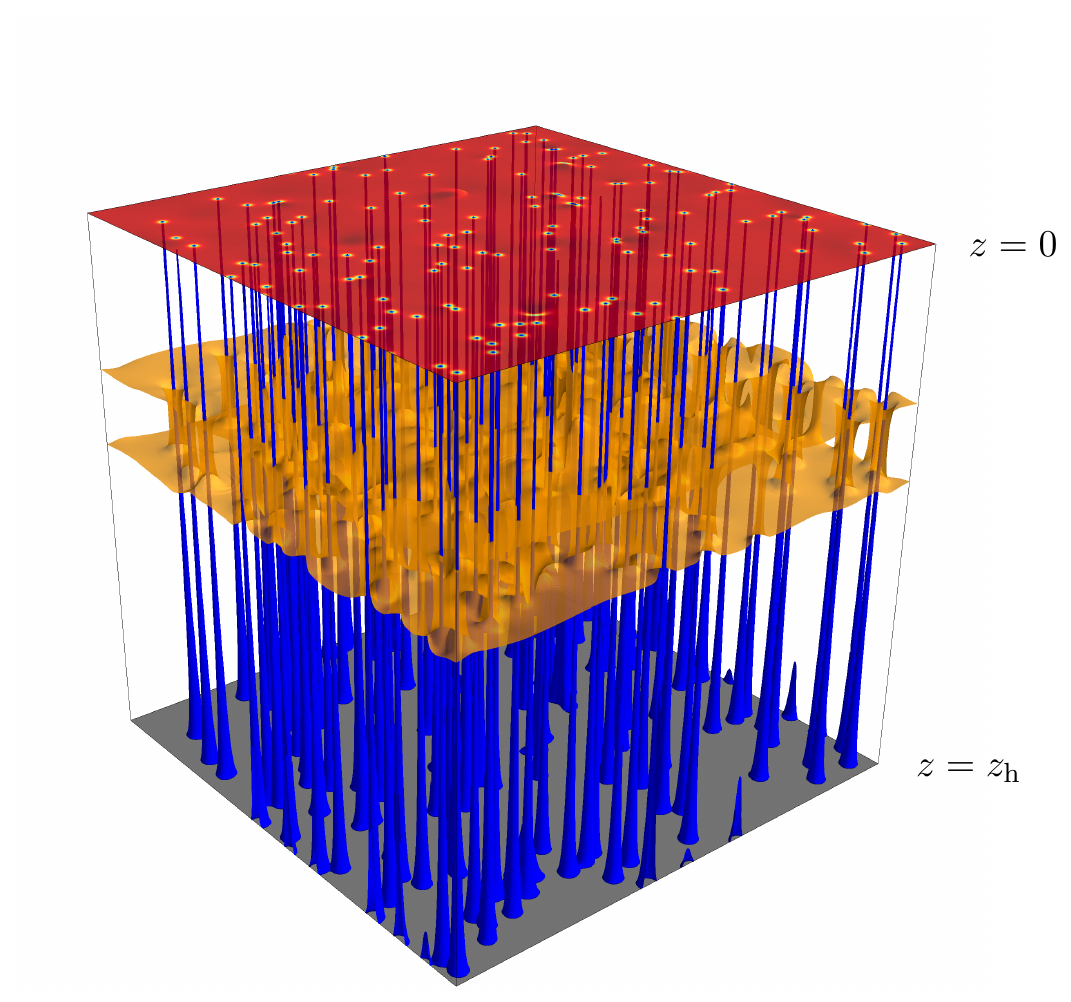}
  \includegraphics[width=0.62\textwidth]{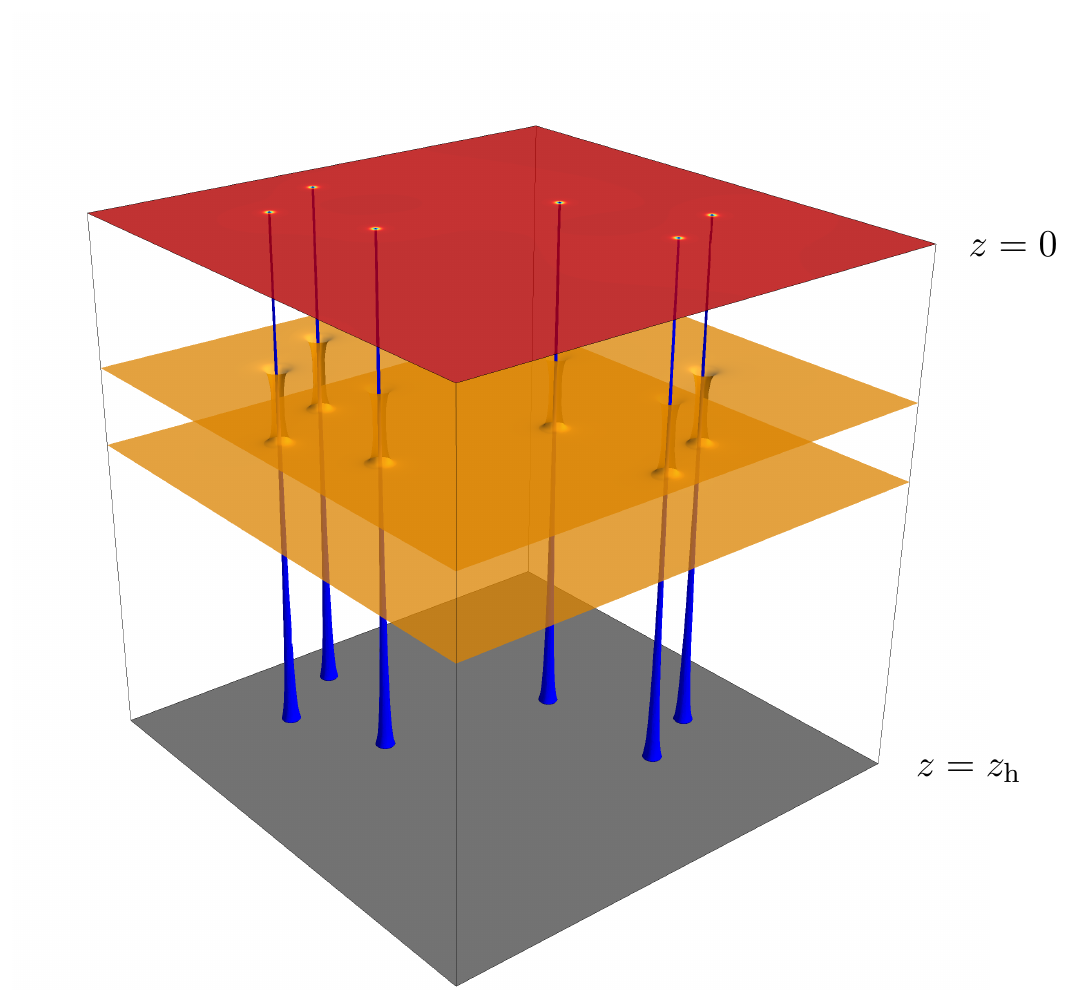}
  \caption{%
    Single-run bulk snapshots of isosurfaces of
    $\left\lvert\Phi\right\rvert^2/z^4$ (blue surfaces, defining value
    $2.2$) and the charge density $\sqrt{-g} \left\lvert
      J^0\right\rvert$ (orange surfaces, defining value $12.3$) after
    a quench of type $\mathcal{A}$\,III (random distribution with 720
    elementary vortices of either sign), at the times $t=200$ (upper
    panel) and $t=4000$ (lower panel). %
    At the boundary, $\left\lvert\Phi\right\rvert^2/z^4$ reduces to
    the superfluid density $n =
    \left\lvert\langle\psi\rangle\right\rvert^2$, and we plot $n$ in
    the top slice using the same colour coding as in
    \Fig{dens-evol}. %
    The plots represent our full simulation domain. Consequently, the
    vortex tubes in the upper panel that appear to be `cut' actually
    close on the opposite side due to periodicity. %
    For an unobstructed view on the isosurfaces of the charge density
    see \Fig{bulk-charge-200-4000}.%
    \label{fig:bulk-view-200-4000}
  }
\end{figure}

Typical bulk views of aspects of the field configuration in the vortex
liquid are shown in \Fig{bulk-view-200-4000}. The upper panel
corresponds to an early to intermediate time $t=200$, and the lower
panel to a late time $t=4000$. We plot isosurfaces of
$\left\lvert\Phi\right\rvert^2/z^4$ (blue surfaces) and of the bulk
charge density $\sqrt{-g}\left\lvert J^0\right\rvert$ (orange
surfaces), see \eqref{eq:11} for the definition of $J^0$. The former
quantity reduces to the superfluid density $n =
\left\lvert\langle\psi\rangle\right\rvert^2$ at the boundary, \ie, as
$z\to 0$. The slice at $z=0$ shows $n$ with the same colour map as in
\Fig{dens-evol}.
The vortices in the boundary theory, discernible as dips in the
superfluid density, are represented by `tubes' in the bulk, with the
bulk scalar $\Phi$ vanishing in their centres. The tubes punch holes
through the charge cloud that hovers in the bulk, and thus provide an
avenue for energy to dissipate into the black hole
\cite{Adams:2012pj}.
A larger defining value of the isosurface of the charge density would
lead to an increased width of the holes and a decreased distance
between the two sheets of the isosurface.
We observe that sound waves in the boundary theory, such as those
produced in vortex annihilation events, are reflected in the bulk as
perturbations of the charge density.
In particular, the isosurfaces of the bulk charge density at $t=4000$
(lower panel), when the system has evolved to a dilute vortex gas, are
much smoother than those at $t=200$ (upper panel), when the superfluid
still exhibits many vortices and a high rate of annihilation events.

\begin{figure}[!t]
  \centering
  \includegraphics{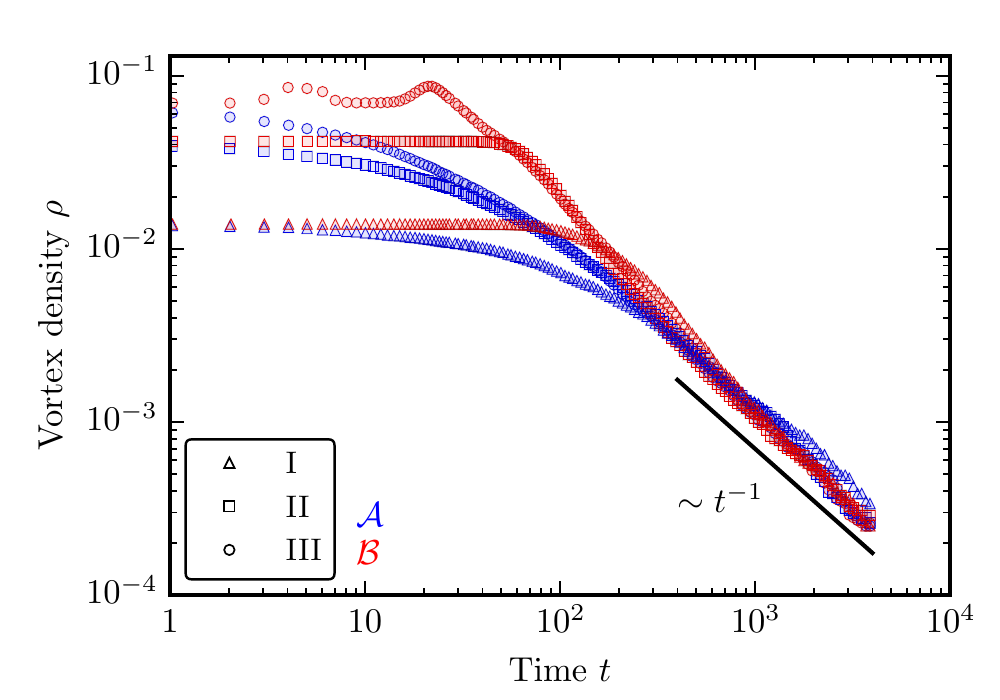}
  \caption{%
    Time dependence of the total vortex density, on a
    double-logarithmic scale. Shown is the averaged vortex density
    $\rho$ for runs starting from the six different initial conditions
    summarised in \Tab{init-conds}. %
    Class-$\mathcal{A}$ (class-$\mathcal{B}$) runs are drawn in blue
    (red), while the different choices of total vortex numbers are
    indicated by different symbols. %
    Note that the apparent initial oscillation of $\rho$ for initial
    condition $\mathcal{B}$\,III is not physical, but rather due to
    uncertainty in the vortex finding algorithm at the high initial
    densities associated with $\mathcal{B}$\,III.%
    \label{fig:decay-vort-dens}
  }
\end{figure}

\begin{figure}[!t]
  \centering
  \includegraphics{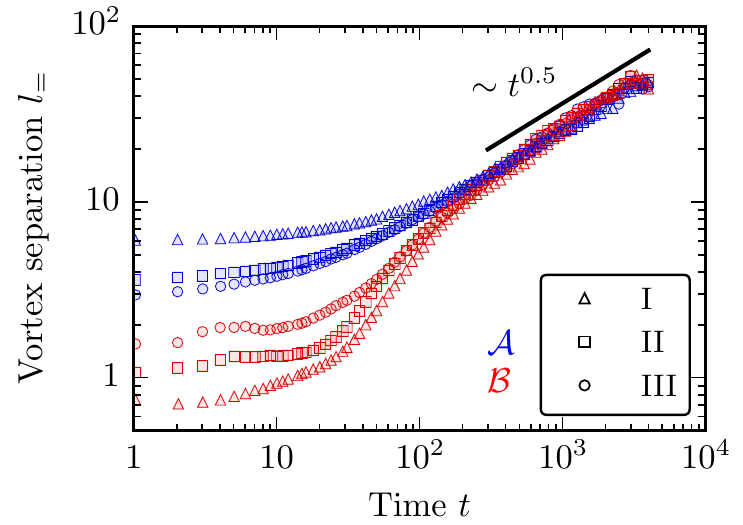}%
  \includegraphics{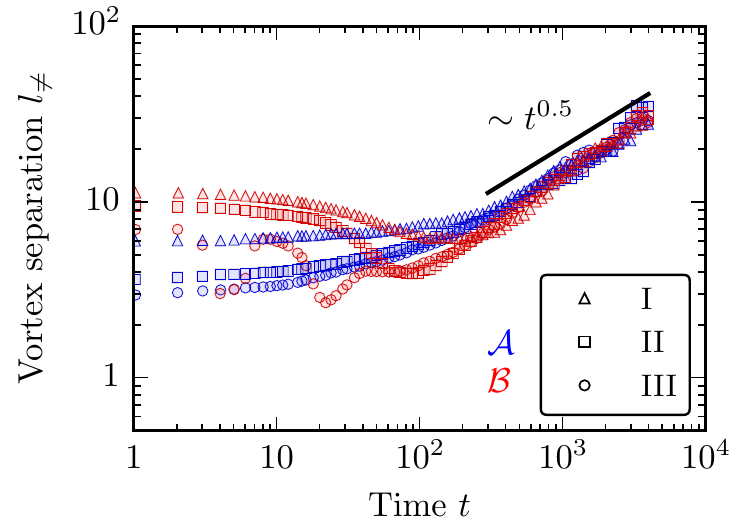}
  \caption{%
    Time evolution of the mean nearest-neighbour vortex--vortex
    distance $\lvv$ (left) and vortex--anti-vortex distance
    $\lva$ (right), in double-logarithmic scale. The class of
    initial condition, $\mathcal{A}$ or $\mathcal{B}$, is colour-coded,
    while the different choices of total vortex number are indicated
    by different symbols, \cf~\Tab{init-conds}.%
    \label{fig:vort-dists}
  }
\end{figure}

Let us now turn to the details of the vortex distribution.
At each unit timestep, we determine the position of all vortices and
anti-vortices, see \Appendix{NumMeth} for details.
In \Fig{decay-vort-dens} we show the total vortex density as a
function of time (averages are taken over ten runs for each type of
quench).
%
%
For both classes of initial conditions, annihilation proceeds in a non-universal manner,
up to a simulation time of $t\simeq 400$, depending on the particular initial configuration.
For initial conditions in class $\mathcal{B}$ (lattice), the vortex density is approximately constant at early times.
This stage was  referred to as the `drift stage' above.
It arises because the like-sign vortex clusters need to expand and dissolve before vortices of opposite
sign can encounter each other. Naturally, this takes longer
if the initial number of vortices is lower.%
\footnote{From our data on the vortex density shown in
  \Fig{decay-vort-dens} and the average vortex separations shown in
  \Fig{vort-dists} below we can estimate the drift velocity $v$ of the
  elementary vortices in the initial stage in which the vortex lattice
  dissolves. We find $v = 0.07, 0.13$, and $0.24$, for the quench types
  $\mathcal{B}$\,I, $\mathcal{B}$\,II, and $\mathcal{B}$\,III,
  respectively.}
At time $t\simeq 400$, starting from any of the initial conditions, a
scaling regime is entered, where the vortex density $\rho$ decays
algebraically in time, $\rho\sim t^{-1}$. This scaling
persists until the end of our simulations at $\tf= 4000$.
We observe that the power-law is the same for all tested initial
conditions.
Hence, starting at $t\simeq 400$, the vortex density has a universal
form. We point out that this decay of vortex density does not exhibit
any characteristic time scale.
Remarkably, the algebraic decay of the vortex density is characterised
by a universal pre-factor. This, together with the value of the
exponent, can be explained by a diffusive motion of vortices within a
thermal background of sound waves.

To investigate this in more detail we define the following quantities.
By $\lvv$ we denote the mean distance of nearest-neighbour defects of equal sign,
while  $\lva$ is the mean nearest-neighbour distance between vortices and anti-vortices,%
\footnote{We note that in both $\lvv$ and $\lva$, the two sums
  corresponding to $\alpha=(+,-)$ need not coincide. However, in the
  thermodynamic limit, we expect that these two sums coincide because
  of the symmetry under exchange of vortices and anti-vortices. We
  have numerically checked that, on average, the two sums are indeed
  equal.}
\begin{align}
  \lvv &= \frac{1}{2}\sum_{\alpha=(+,-)}\frac{1}{N} \sum_{i_\alpha=1}^N \left\lvert\vec{x}_{i_\alpha}-\vec{x}_{\text{nn}_\alpha(i_\alpha)}\right\rvert \,,\label{eq:13}\\
  \lva &= \frac{1}{2}\sum_{\alpha=(+,-)}\frac{1}{N} \sum_{i_\alpha=1}^N \left\lvert\vec{x}_{i_\alpha}-\vec{x}_{\text{nn}_{-\alpha}(i_\alpha)}\right\rvert \,.\label{eq:14}
\end{align}
Here, $\alpha=(+,-)$ denotes the sign of the winding number of the
vortex indexed by $i_\alpha=1,\dots,N$, and $\vec{x}_{i_\alpha}$ is
its position.  The function
$\text{nn}_\alpha(i_\beta)$ yields the index of the vortex of sign
$\alpha$ nearest to the vortex $i_\beta$.

In \Fig{vort-dists} we show  $\lvv$ (left panel) and  $\lva$ (right panel) as functions of
time (averages are taken over ten runs for each type of
quench).
%
%
As the inverse of the square root of the vortex density gives the mean
vortex distance, the time evolution of the above length scales
is directly related to the density evolution shown in \Fig{decay-vort-dens}.
The panels in \Fig{vort-dists} show that the nearest-neighbour
distances evolve as $\lvv(t) = \sqrt{4\Dvv\, t}$ and $\lva(t) =
\sqrt{4\Dva\, t}$ during the universal late-time period, \ie,
diffusively.
From our data shown in \Fig{vort-dists}, we estimate $\Dvv \approx
0.16$ and $\Dva \approx 0.05$.
The diffusive behaviour is a consequence of the effective interaction
of vortices with the thermal background in the dual theory.  We
attribute the fact that $\Dvv$ and $\Dva$ differ to the
influence of the effective vortex--vortex and vortex--anti-vortex interactions.
The values of the constants $\Dvv$ and $\Dva$ can be used to further
constrain the parameters controlling the dissipation and the vortex
interactions of the boundary theory for a quantitative comparison with
conventional models of superfluidity.
Finally, note that the universal regime associated with this
scaling behaviour of both $\lvv$ and $\lva$ is entered at the same time $t\simeq 400$
for all initial conditions.

\subsection{Holographic turbulence}
\label{sec:critical-dynamics}
So far we have identified at late times $t\gtrsim 400$ a universal
stage of the system's evolution with respect to the effective dynamics
of vortices.
During this stage, the characteristic length scales of the
vortex distribution evolve algebraically, as we have shown in
\Sect{stag-dynam-evol}. Therefore, their rate of change, $\dot{l}/l$,
decreases and approaches zero asymptotically. Thus, at late times the
system can be considered quasi-stationary.
As we will discuss in the following, during this stage the system bears signatures of turbulence.
We will, in particular, analyse momentum-space distributions as statistical observables and find them to exhibit algebraic behaviour as it is characteristic for the development of turbulent transport.

While vortices are the building blocks of superfluid turbulence, one
needs to study correlation functions of the superfluid order-parameter
field in order to obtain a full understanding of the microscopic
dynamics of the superfluid.
Here, we concentrate on the
equal-time two-point correlation function,
$\langle\psi^{\ast}(\vec{x},t)\psi(\vec{y},t)\rangle$. Since our
system is spatially homogeneous in a statistical sense this quantity
can be analysed in relative momentum space, \ie~with respect to the
relative coordinate $\vec{r} = \vec{x} - \vec{y}$.
One defines what is known as the (radial) occupation number spectrum
as%
\footnote{\label{fn:ensavrgs}A comment is in order here. In practice,
  we extract the full quantum expectation value
  $\langle\psi(\vec{k},t)\rangle$ from the simulations, and compute
  $\int\langle\psi^*(\vec{k},t)\rangle\langle\psi(\vec{k},t)\rangle
  \,\D\Omega_k/(2\pi)$ from it as an approximation for $n(k,t)$.
  In addition, we average this quantity over a number of 
  runs \cite{Chesler:2014gya}.  The infrared phenomena
  we are interested in here are expected to be classical in the
  statistical sense, and averaging over many realisations washes
  out, e.g., effects due to the definite positions of the vortices.
  It would be nonetheless interesting, if numerically
  demanding, to use the holographic dictionary to extract the full
  quantum two-point function.}
\begin{equation}
  \label{eq:4}
  n(k,t) = \int\!\frac{\D\Omega_k}{2\pi}\, \langle\psi^*(\vec{k},t)\psi(\vec{k},t)\rangle \,.
\end{equation}
Due to the isotropy of the underlying model we can perform the angular
integration without losing information, such that we are left with the
radial momentum $k=\lvert\vec{k}\rvert$.
The standard radial kinetic energy spectrum is defined as
\begin{equation}
  \label{eq:rad-kin-en}
  E(k) = \int \frac{\D\Omega_k\,k}{2\pi} \lvert\vec{k}\rvert^2\langle\psi^*(\vec{k},t)\psi(\vec{k},t)\rangle
\end{equation}
and is related to the radial occupation number spectrum via $k^3 n(k)
= E(k)$.
In general $n(k)$ has a well-defined field-theoretic interpretation.

In his seminal work on turbulence
(\cite{Kolmogorov1941a,Kolmogorov1941b,Kolmogorov1941c}, see also
\cite{Frisch2004a}), Kolmogorov assumed the existence of an inertial
range in momentum space, bounded by two characteristic scales,
$k_{\text{in}}$ and $k_{\text{diss}}$.
At the scale $k_{\text{in}}$, energy is injected into the system. It
`cascades' from momentum shell to momentum shell before it is eventually
dissipated into heat at the higher scale $k_{\text{diss}}$. Based on the
assumption of such a local transport in momentum space,
Kolmogorov found that the kinetic energy spectrum of stationary
turbulent flow in an incompressible fluid
scales as $E(k)\sim k^{-5/3}$ within the inertial range.
This result holds also in two spatial dimensions where it is
associated with an inverse energy cascade \cite{Kraichnan1967a}.
The essential feature of this so-called Kolmogorov--Richardson cascade
is that the system is self-similar, \ie~scale-free, within the inertial range.

Here, we take a more general view on turbulence and analyse the
occupation number spectra of the superfluid in terms of scaling laws,
$n(k)\sim k^{-\zeta}$, previously discussed in \cite{Adams:2012pj}.
We extract the scaling exponents $\zeta$ by fitting power laws to the
spectra after averaging them over ten runs for each initial
condition. We estimate the uncertainty of the fitted exponents to be
$0.1$, for details see \Appendix{determ-scal-law}.
In the following, we relate the scaling seen in the spectra to the
statistical properties of the vortex distribution that we discussed in
\Sect{stag-dynam-evol}. In particular, we want to study whether the
universality that emerges in the late-time vortex dynamics is
reflected in the occupation number spectra.

\begin{figure}[!t]
  \centering
  \includegraphics{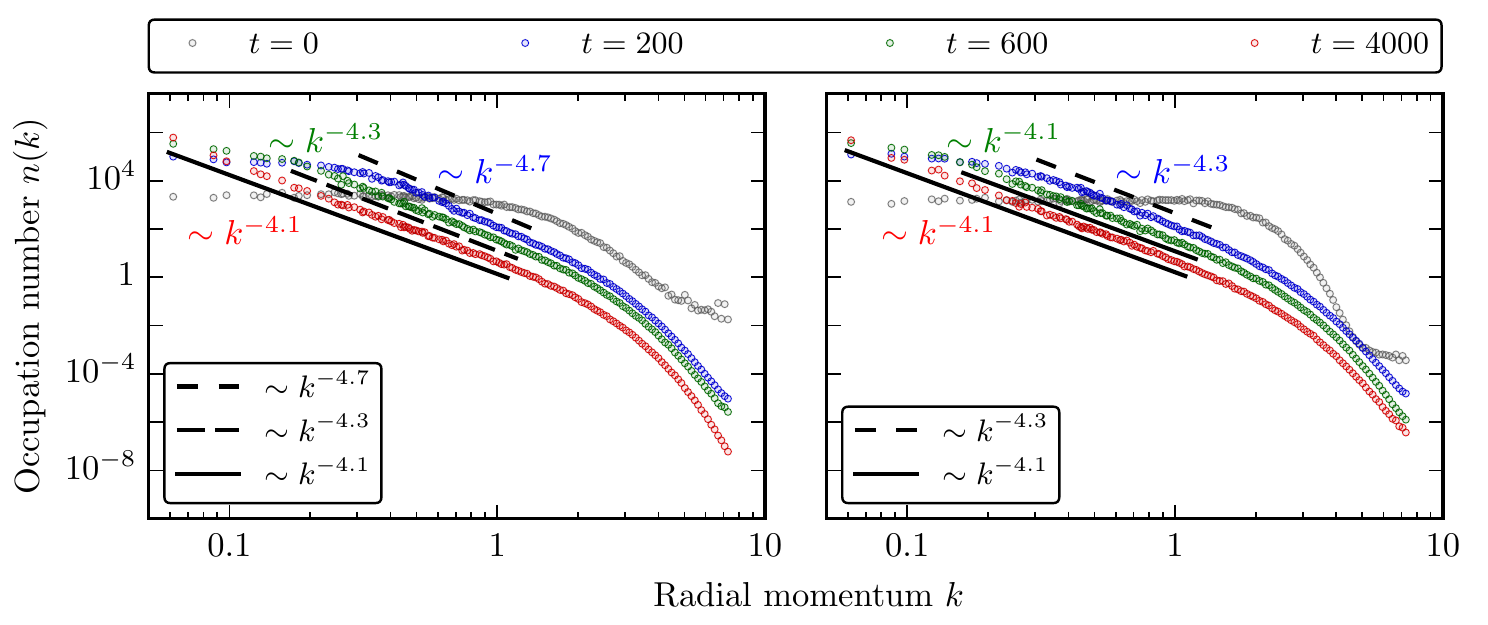}
  \caption{%
    Occupation number spectra as functions of the radial momentum $k$
    during the different stages of the evolution, on a
    double-logarithmic scale. Here we show the evolution after quench
    type $\mathcal{A}$\,II (random distribution, left), and the
    evolution after quench type $\mathcal{B}$\,II (vortex lattice,
    right). %
    Results for the respective parameter sets I and III are similar to
    the results depicted here, \cf~the main text.%
    \label{fig:number-specs-432}
  }
\end{figure}

In \Fig{number-specs-432}, we show the occupation number spectra
$n(k)$ for the initial conditions $\mathcal{A}$\,II and
$\mathcal{B}$\,II (averages are taken over ten runs for each type of
quench). To illustrate
their time evolution the spectra are shown at $t=0, 200, 600$, and
4000.
In both classes, $\mathcal{A}$ (random distributions) and
$\mathcal{B}$ (vortex lattices), the spectra for the parameter sets I
and III are very similar to the spectra for the sets II of the
respective class.
For intermediate and late times we observe inertial ranges in $k$, the
corresponding power laws fitted to the spectra at these times are also
shown in the figure. Within the uncertainty in our determination of
scaling exponents of occupation spectra
(\cf~\Appendix{determ-scal-law}), the scaling exponents at intermediate to late times, $t\gtrsim 400$,
agree even quantitatively for all quench types within each
class. Thus, we can discuss the generic features of the spectra at the
examples shown in \Fig{number-specs-432}.
%

First, we note that the initial occupation spectra (corresponding to
time $t=0$ in \Fig{number-specs-432}) are indeed similar to the
far-from-equilibrium initial momentum distributions considered in
various classical statistical simulations of dilute Bose gases
\cite{Nowak:2010tm,Nowak:2011sk,Schole:2012kt,Karl:2013mn,Karl:2013kua}.
Similar initial momentum distributions have also been studied
in the relativistic case  \cite{Berges:2010ez,Berges:2012us}.

During the early to intermediate stage of the system's evolution, for example
at $t=200$, the scaling exponents differ between the various
initial conditions, indicating that this stage is still
non-universal. This could be expected from our analysis of the vortex
dynamics in \Sect{stag-dynam-evol} where we saw that a dependence on
the initial parameters persists until $t\simeq 400$.
Specifically, we find, at time $t = 200$, a scaling at intermediate
momenta with exponents $\zeta\approx 4.7$ and $\zeta\approx 4.3$ in
class $\mathcal{A}$ and $\mathcal{B}$, respectively, for quench types II and III.
The exponents for quench type I in each of the classes
$\mathcal{A}$ and $\mathcal{B}$ are slightly larger, $\zeta\approx
4.9$ and $\zeta\approx 4.5$, respectively.
The power-law scaling $n(k) \sim k^{-4.7}$ of the occupation number $n(k)= k^{-3}E(k)$ at
intermediate times after quenches $\mathcal{A}$\,II and $\mathcal{A}$\,III (random
distributions) is consistent with an $E(k)\sim k^{-5/3}$ Kolmogorov scaling of
the radial kinetic energy spectrum.
The scaling exponent found in the case of initial condition
$\mathcal{A}$\,I is slightly larger than the Kolmogorov value.

At this point, let us briefly pause to compare our findings for the intermediate times (at about $200 \lesssim t \lesssim 400$) to the
results reported in \cite{Adams:2012pj}. There, the evolution of the
holographic superfluid was studied starting from an initial condition
very similar to our type $\mathcal{B}$\,II vortex-lattice
configuration. Even though the simulation domain used in
\cite{Adams:2012pj} was smaller than the one employed here, the
initial vortex densities differ by only $10\,\%$.
The kinetic energy spectrum was defined in \cite{Adams:2012pj}, based
on hydrodynamic arguments \cite{Nore1997b}, as
${\epsilon_\mathrm{kin}}(k,t) = \frac{1}{2}\int
\left\lvert\vec{w}(\vec{k},t)\right\rvert^2 k\,\D\Omega_k$. Here,
$\vec{w}(\vec{k},t)$ is the Fourier transform of the generalised
velocity field $\vec{w}(\vec{x},t) = \langle\psi(\vec{x})\rangle
\vec{\nabla}\varphi(\vec{x})$ with $\varphi=\arg(\langle\psi\rangle)$ the
phase of the superfluid order parameter.
This definition of the kinetic energy spectrum differs from that of
our $E(k)$ in \Eq{rad-kin-en}.
In the cited work, the system was propagated up to time $t=600$, and
Kolmogorov scaling of the kinetic energy spectrum,
$\epsilon_{\text{kin}}(k,t)\sim k^{-5/3}$, was reported for
intermediate evolution times in the range $160 < t < 500$.
We have analysed our data for quench type $\mathcal{B}$\,II also in
terms of the quantity $\epsilon_{\text{kin}}(k,t)$.
We find, within the uncertainty inherent in the fitting procedure, a
scaling law $\epsilon_{\text{kin}}(k,t)\sim k^{-1.3}$ during the
intermediate stage of the evolution, for example at $t=200$, differing
from Kolmogorov scaling.

We now continue with the discussion of our results in \Fig{number-specs-432}.
During the evolution, the momentum range where
$n(k,t)$ obeys a scaling law gradually grows on its lower end, for all
initial conditions.
Also, the scaling exponents' absolute values
decrease. This can be attributed to the increasing diluteness of the
vortex gas.
The flow field of a single vortex exhibits a scaling $n(k) \sim
k^{-4}$ for momenta not resolving the vortex core. The same scaling is
expected for randomly distributed vortices and anti-vortices at
momenta larger than the mean inverse of the vortex--anti-vortex
distance \cite{Nowak:2011sk}.
Hence, the inverse average vortex separation sets a lower cutoff to
the scaling regime. As the vortex gas becomes more dilute, the average
vortex separation increases, see \Fig{vort-dists}, so that the lower
cutoff of the scaling regime decreases. Furthermore,
\Fig{number-specs-432} shows that the scaling exponents gradually
approach the value for single-vortex scaling.

\begin{figure}[!t]
  \centering
  \includegraphics{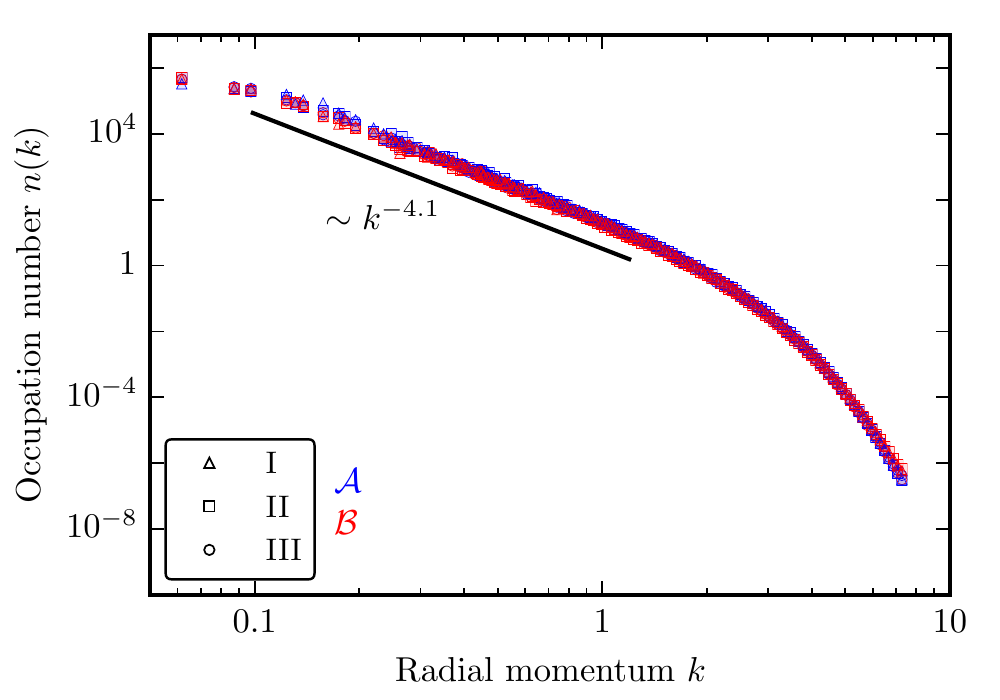}%
  \caption{%
    Occupation number spectrum as a function of the radial momentum
    $k$ at time $t=1000$ during the universal late stage of the evolution,
    on a double-logarithmic scale. The class of quenches, $\mathcal{A}$
    or $\mathcal{B}$, is colour-coded, while the different choices of
    total vortex number are indicated by different symbols,
    \cf~\Tab{init-conds}. %
    The spectra exhibit scaling $n(k)\sim k^{-\zeta}$ with
    $\zeta\approx 4.1$ over a large momentum range that gradually
    expands to the infrared.%
    \label{fig:number-specs-all}
  }
\end{figure}

Let us next turn to the late-time evolution.  At time $t=600$, the
scaling exponents of the occupation number spectra are still slightly
different for quench classes $\mathcal{A}$ and $\mathcal{B}$. They are
fitted by $\zeta\approx 4.3$ and $\zeta\approx 4.1$ for quench types
$\mathcal{A}$\,II and $\mathcal{B}$\,II, respectively.
At time $t=4000$ the scaling exponent is fitted by $\zeta\approx 4.1$ for
both quench types $\mathcal{A}$\,II and $\mathcal{B}$\,II.
We recall that the uncertainty in the estimation of these scaling
exponents is about 0.1, see \Appendix{determ-scal-law}.
As discussed above, within each of the classes $\mathcal{A}$ and
$\mathcal{B}$ of quenches, at late times both the qualitative and quantitative
scaling behaviour is unchanged for the alternative choices I and III of
the total vortex number. In particular, the scaling exponents in the
late stages of the evolution are consistent with
$4.1\lesssim\zeta\lesssim 4.3$ for all initial conditions. To
demonstrate this, we plot in \Fig{number-specs-all} the occupation
number spectra for all our quenches (averages are taken over ten runs
for each type of quench) at $t=1000$ during the late-time stage.
At this time, for all initial conditions, the estimates for the
scaling exponents are in fact in the narrower range $4.1\lesssim
\zeta \lesssim 4.2$.
%
%
This indicates the emergence of a universal scaling law for
$n(k,t)$. Thus, in the long run, the system loses memory of its
initial conditions, confirming our findings in \Sect{stag-dynam-evol}.
Furthermore, the universal scaling law in the occupation number
spectra emerges at late times $t\gtrsim 600$. This is to be
compared with our findings in \Sect{stag-dynam-evol}. There, we identified
$t\gtrsim 400$  to roughly mark the onset of the late-time
universal scaling behaviour with respect to time.

We note that the late-time scaling exponent still appears to deviate slightly
from the single-vortex scaling $\zeta=4$, indicating that effects of
vortex interactions are still relevant in this universal regime.
Typically, for all initial conditions, we still have 4 to 8 remaining
vortex defects at the final time $\tf= 4000$.

\section{Non-thermal fixed point: the holographic perspective}
\label{sec:hNTFP}
In the following, we concentrate on the universal late-time behaviour
of the holographic superfluid.
As we have found in the previous section, the system exhibits two
types of universal scaling laws, one in space and one in time, at late
times $t\gtrsim 600$.
The first type characterises spatial correlations of excitations and
is reflected by the scaling $n(k) \sim k^{-\zeta}$ of the
single-particle occupation number spectrum with the exponent in the
range $4.1\lesssim \zeta\lesssim 4.3$.
This scaling is observed within an infrared momentum regime
which can be viewed as an inertial range. The
power-law behaviour terminates in the infrared at a momentum scale
that is related to the characteristic length scales of
the vortex distribution.
The characteristic lengths of the vortex distribution, including the
mean vortex separation, follow a scaling law in time, $l \sim
t^{1/2}$, see \Sect{stag-dynam-evol}.
This leads to the observation that the system's time evolution
is slowing down algebraically during the universal stage, $\dot l / l
\sim t^{-1}$.
Such a behaviour is interpreted as `critical slowing down' within the
context of dynamic critical phenomena well-known from the theory of
dynamics near an equilibrium phase transition \cite{Hohenberg1977a}.

Similar scaling features have been observed in numerical calculations
of dilute Bose gases far from equilibrium on the basis of a statistical evaluation
of Gross--Pitaevskii models
\cite{Nowak:2011sk,Schole:2012kt,Nowak:2012gd}. Within
that framework, the scaling features have been interpreted in terms of
the more general concept of non-thermal fixed points
\cite{Berges:2008wm,Berges:2008sr,Scheppach:2009wu}. This concept associates
stationary points in the time evolution of non-thermally scaling correlation functions with
fixed points in a renormalisation-group sense
\cite{Gasenzer:2008zz,Berges:2008sr,Mathey2014a.arXiv1405.7652}. In the vicinity of those fixed points,
correlation functions assume spatial and temporal scaling
behaviour.
Non-thermal fixed points thus constitute a generalisation of the
concept of critical phenomena near thermal equilibrium.
A priori, this situation allows for new universality classes
as compared to those in the classification of Hohenberg and Halperin
\cite{Hohenberg1977a}.
But also phase-ordering kinetics as discussed in \cite{bray1994}
within the Hohenberg--Halperin classification scheme could probably be
interpreted as a realisation of a non-thermal fixed point.

\begin{figure}[!t]
  \centering
  \includegraphics{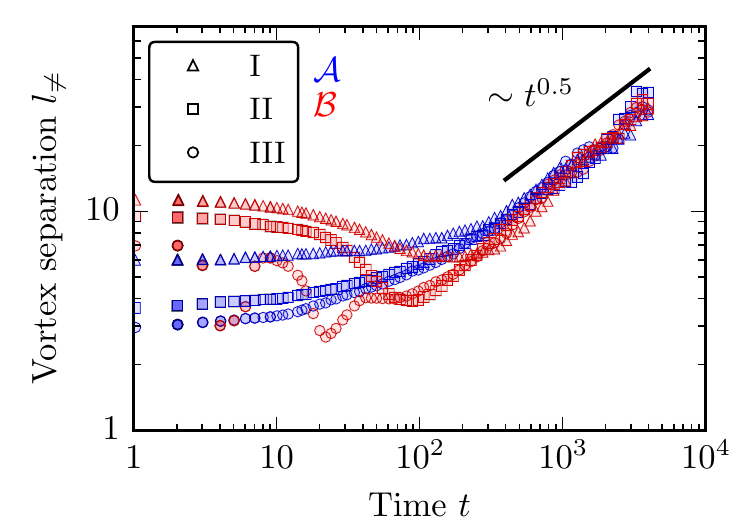}%
  \includegraphics{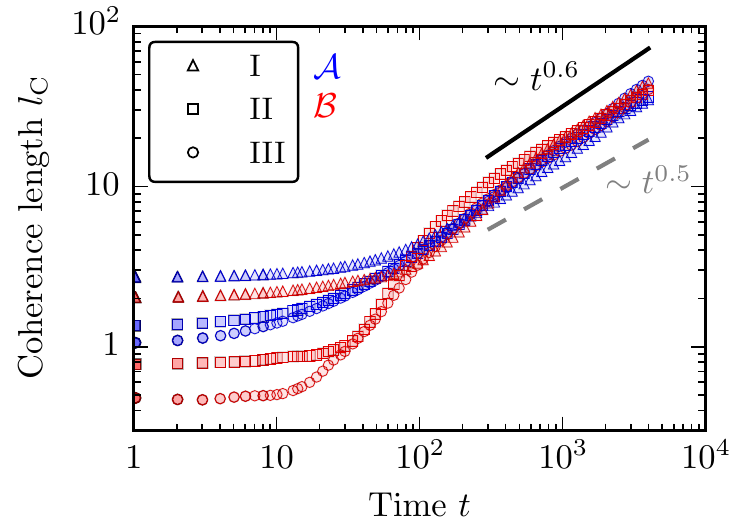}
  \caption{%
    Nearest-neighbour vortex--anti-vortex distance $\lva$ (left panel)
    and coherence length $l_{\text{C}}$ (right panel) as functions of
    time for all initial conditions. The type of initial condition,
    $\mathcal{A}$ or $\mathcal{B}$, is colour-coded, while the
    different choices of total vortex number are indicated by
    different symbols, \cf~\Tab{init-conds}.%
    \label{fig:lD-lCstar}
  }
\end{figure}

In the following, we argue that the holographic superfluid indeed
exhibits the presence of a non-thermal fixed point. We use observables which have been introduced in \Cite{Schole:2012kt} for the purpose
of numerically identifying non-thermal fixed points in two-dimensional
Bose gases.
The state of the system can be characterised by two length scales, the
mean nearest-neighbour vortex--anti-vortex distance $\lva$, see
\Eq{14}, and the coherence length $l_\text{C}$ of the superfluid.
The latter is defined as
\begin{equation}
  \label{eq:def-lC}
  l_\text{C}(t) = \int g^{(1)}(r,t)\,\D r
\end{equation}
with the autocorrelation function $g^{(1)}$,%
\footnote{Concerning the expectation value involved in the computation
  of $g^{(1)}$ cf.~our remarks in footnote \ref{fn:ensavrgs}.}
\begin{equation}
  g^{(1)}(r,t) = \int\frac{\D^2x}{A}
  \int\frac{\D\Omega_r}{2\pi}\,
  \frac{\langle\psi^*(\vec{x},t)\psi(\vec{x}+\vec{r},t)\rangle}{\sqrt{n(\vec{x},t)
      n(\vec{x}+\vec{r},t)}} \,.
\end{equation}
Here, $\int\D^2x/A$ denotes the average over the simulation domain $A$, and $\int\D\Omega_r/(2\pi)$ the
angular average for the difference vector $\vec{r}$.
$\lva$ is a measure for the state of the system from the point of view of the effective vortex
picture. The coherence length $l_\text{C}$
measures the degree of spatial correlations of microscopic
excitations. The time evolution of both lengths is depicted in
\Fig{lD-lCstar} (averages are taken over ten runs for each quench
type). Both $\lva$ and $l_{\text{C}}$ show universal
scaling behaviour in the late stage of the evolution. The
scaling exponent of $\lva$ is found to be $0.5$, as discussed
in \Sect{stag-dynam-evol}.
The scaling exponent of $l_\text{C}$ can be narrowed down to a value
between $0.5$ and $0.6$.
Thus, it is conceivable that the latter scaling exponent agrees with
the scaling exponent of the characteristic length scales of the vortex
distribution. This would imply a constancy of ratios of different
characteristic length scales typical for the approach to a non-thermal
fixed point or a critical point, see for example \cite{Berges:2004ce}.
%

%
For the temperature and chemical potential we have chosen, the scaling
exponents that we have extracted happen to coincide with those
predicted by the theory of phase-ordering kinetics \cite{bray1994}.
Note that, as the holographic superfluid is a strongly correlated
quantum fluid, it is a non-trivial question to what extent the
arguments of \cite{bray1994} apply.  In addition, going beyond the
probe limit could alter these findings.

\begin{figure}[!t]
  \centering
  \includegraphics{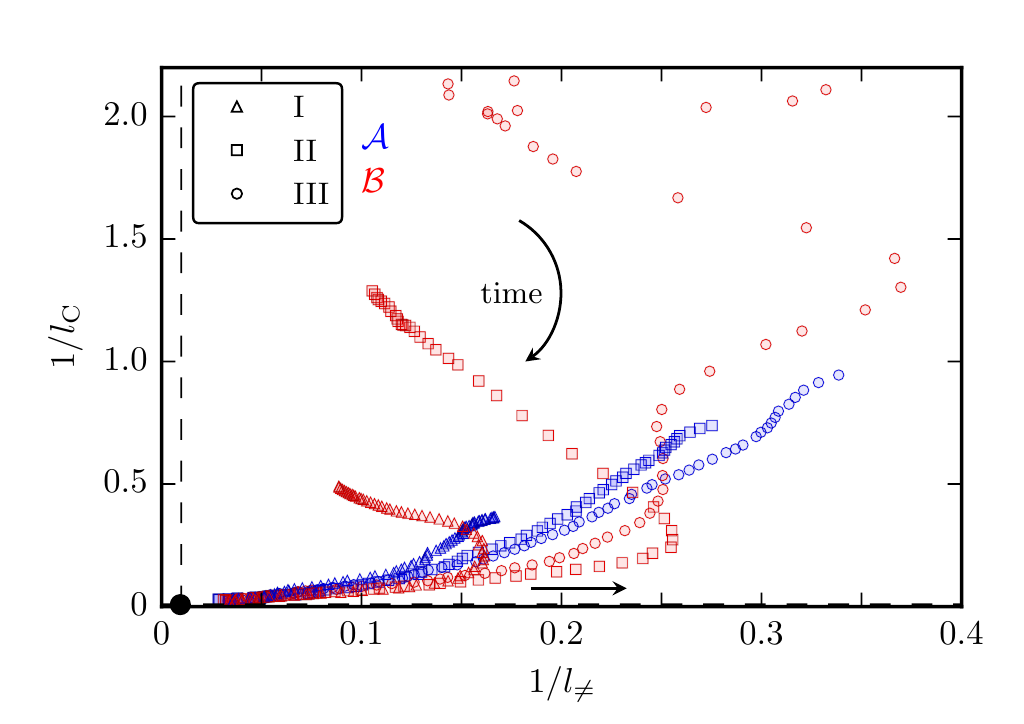}
  \caption{%
    Approach of the non-thermal fixed point:  Trajectories of the system
    in the plane spanned by the inverse coherence length $1/l_{\text{C}}$ and
    the inverse mean nearest-neighbour vortex--anti-vortex distance
    $1/\lva$. The type of initial condition, $\mathcal{A}$ (random
    distribution) or $\mathcal{B}$ (vortex lattice), is colour-coded,
    while the different choices of total vortex number are indicated
    by different symbols, \cf~\Tab{init-conds}. %
    The points are approximately equidistant on a logarithmic time
    scale: Given the time $t_n$, the time of the next point is given
    by $t_{n+1}=1.1\, t_n$. The first point of each curve represents
    $t_1=1$. %
    The dashed lines correspond to the largest coherence length and
    vortex--anti-vortex distance possible on the finite simulation
    domain. %
    For time $t\to\infty$, all curves would bend over towards
    $(1/\lva, 1/l_{\text{C}}) \to (\infty, 0)$, as indicated by an
    arrow.%
    \label{fig:flow-plot}
  }
\end{figure}

It is useful to study the time evolution of the system in a reduced
configuration space consisting of the two scales $\lva$ and
$l_{\text{C}}$.
\Fig{flow-plot} shows the trajectories of the system for different
initial conditions (averages are taken over ten runs for each quench
type) in this configuration space.
Here we choose to plot the inverse lengths, $1/\lva$ and
$1/l_{\text{C}}$, on the axes because in the thermodynamic limit one
generically expects characteristic length scales to diverge at
critical points.
While the inverse coherence length decreases monotonically for all
initial conditions, the behaviour of $1/\lva$ is different for the two
classes of initial conditions.
Starting from random distributions of vortices (class $\mathcal{A}$),
$1/\lva$ decreases monotonically due to vortex--anti-vortex
annihilation.
For vortex lattices (class $\mathcal{B}$) $1/\lva$ increases during
the initial drift phase as the vortices of opposite sign move closer
to each other.
Subsequently, $1/\lva$ decreases monotonically for all of our initial
conditions during the universal stage.
Eventually, for all of our initial conditions the system is attracted
by the point of maximal coherence and maximal vortex--anti-vortex
separation, as marked in the figure. Close to the fixed point, all
trajectories meet in a universal curve, indicating that the memory of
the initial condition is completely lost and, with that, signalling
the universal stage of time evolution.
Due to the algebraic slowing-down the system spends a major part of
its time evolution near the fixed point.
For time $t\to\infty$, all curves would bend over towards $(1/\lva,
1/l_{\text{C}}) \to (\infty, 0)$ because the last vortex--anti-vortex
pair annihilates and the coherence length diverges.%
\footnote{Obviously, in our simulation this happens only to the extent
  possible on a finite domain.}
This would mark the approach of thermal equilibrium.
%
%
Our findings indicate that the abstract concept of a non-thermal fixed
point in the quantum dynamics can be interpreted in a simple way on
the gravity side.
Specifically, a fixed point corresponds to a time-independent solution
of the classical bulk equations of motion \eqref{eq:1} and
\eqref{eq:8}.
There is one unique static solution which is completely stable and
corresponds to thermal equilibrium, \ie~the thermal fixed point. But,
as for many systems of non-linear partial differential equations,
there will be a series of stationary points with different stability
characteristics.
The intriguing properties of non-thermal fixed points are reproduced
by partially stable stationary points, \ie~points which have at least
one attractive direction in phase space.
This guarantees that the time evolution first approaches such a fixed
point -- regardless of details of the initial condition -- before
turning towards thermal equilibrium.
Here, we have identified such a partially stable stationary point for
the dynamics of the holographic model in the reduced configuration
space of \Fig{flow-plot}.

Let us now discuss the picture developed of the non-thermal fixed
point in the context of vortices in a superfluid
\cite{Nowak:2010tm,Nowak:2011sk,Schole:2012kt}.  Precisely at a
non-thermal fixed point, scaling of, e.g., the momentum distribution
applies by definition at all momenta smaller than some ultraviolet
scale, such as the healing length measuring the distance over which a
density perturbation `heals out' in the superfluid.  A fixed-point
momentum distribution $n(k)\sim k^{-4}$ can be traced to the geometry
of the fluid-velocity field around a vortex core, and extending this
to $k=0$ would imply a single vortex at a random position in an
infinitely extended system.  Hence, vanishing total angular momentum
as in the examples discussed above, would require, approaching the
fixed point, the `vortex-behind-the-moon' scenario where the
opposite-winding-number vortex is at a large distance from the first
one.
   
Considering this idealised picture, the fixed point can never be
exactly reached in a finite system such as on our computer
grid. However, the diluting vortex gas approaches a configuration with
a few far-separated vortices with vanishing total angular momentum, in
fact close to the scenario described above.  Notwithstanding these
limitations also in a finite geometry metastable vortex configurations
exist such as regular lattices of vortices with alternating winding
number within a noise-free condensate field.  From a
renormalisation-group point of view such configurations could
presumably be described in the framework of universal scaling
functions in the presence of an infrared cutoff, requiring a
characterisation of the space of running couplings in the surroundings
of the fixed point.

Finally, we would like to look in more detail at the properties of the
non-thermal fixed point in terms of bulk properties in the holographic
description. As discussed above, the fixed point itself cannot be
reached in our simulations. But it is natural to expect that the
quasi-stationary configurations in the late-time stage of the system's
evolution should in many respects be close to the fixed point.

\begin{figure}[!t]
  \centering
  \includegraphics[width=0.65\textwidth]{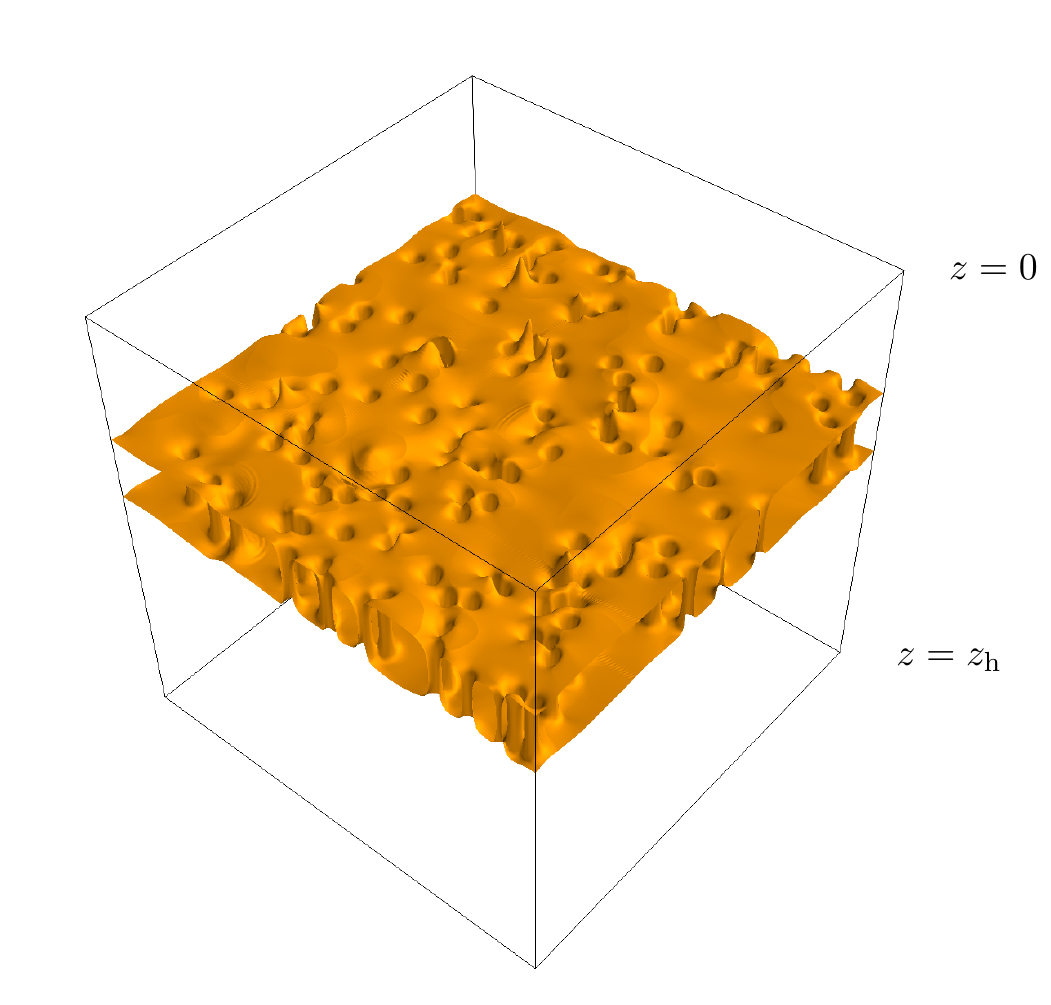}
  \includegraphics[width=0.65\textwidth]{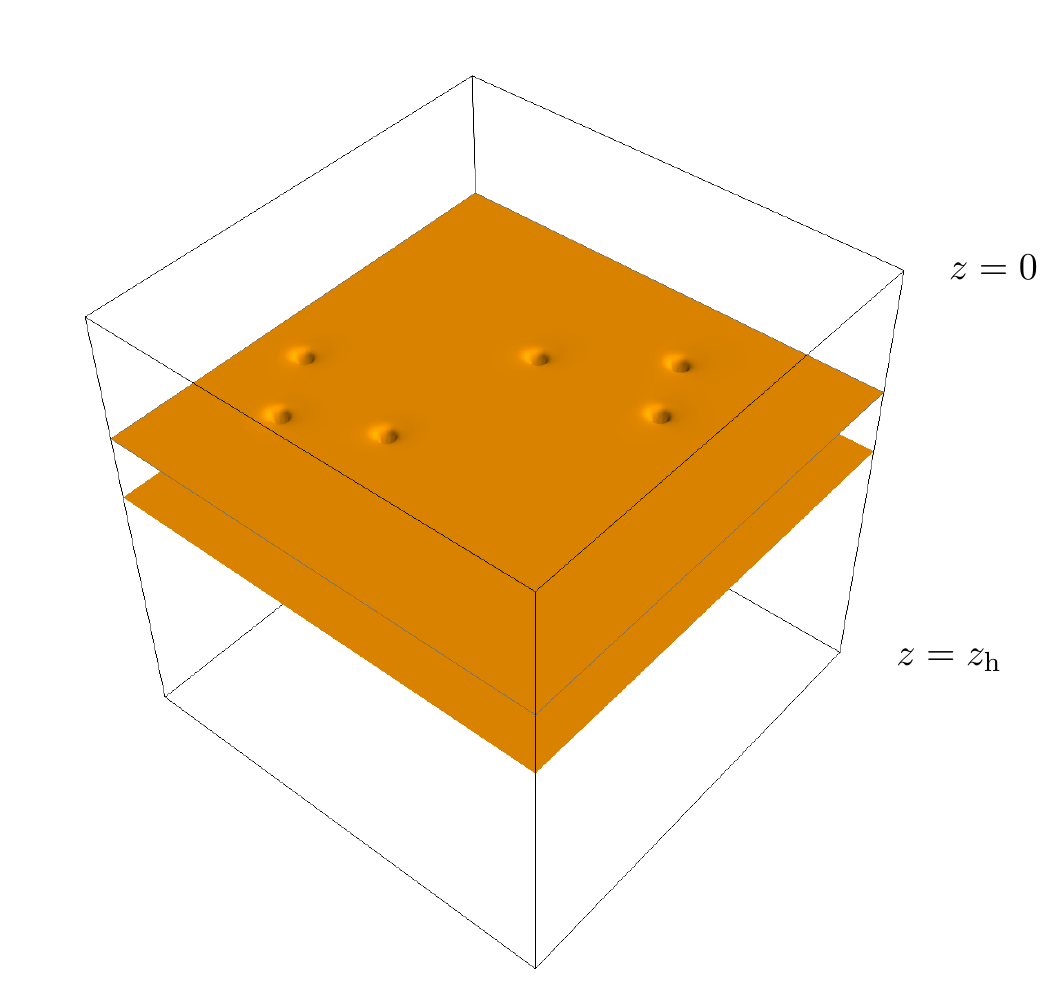}
  \caption{%
    Single-run bulk snapshots of isosurfaces of the charge
    density $\sqrt{-g} \left\lvert J^0\right\rvert$ at times
    $t=200$ (upper panel) and $t=4000$ (lower panel). The same
    parameters as in \Fig{bulk-view-200-4000} are used. %
    The holes in the charge density indicate the presence of
    vortices. %
    Note the strong variations of the charge density field at
    $t=200$. The isosurface exhibits ripples resulting from vortex
    annihilation events. In contrast to this, the isosurface at
    $t=4000$ is very smooth.%
    \label{fig:bulk-charge-200-4000}
  }
\end{figure}

We observe that in the early stage of the evolution the bulk fields
exhibit strong variation in the $x$- and $y$-directions.  In
particular, these variations are also present away from the vortex
cores. In the quasi-stationary late-time regime, in contrast,
significant variations are observed only in the direct vicinity of the
vortex cores. This characteristic difference is nicely seen in the
behaviour of the isosurfaces of the quantity $\sqrt{-g}\left\lvert
  J^0\right\rvert$ that we show in \Fig{bulk-view-200-4000}. To
exhibit this more clearly, we plot in \Fig{bulk-charge-200-4000} the
same charge-density isosurfaces as in \Fig{bulk-view-200-4000}, with
all other features of the plots removed.
In the early stage of the evolution (upper panel), the isosurface of
the bulk charge density shows many small ripples and several spikes
resulting from vortex annihilation events. On the other hand, in the
late-time regime (lower panel) the isosurface appears almost
featureless except for the holes due to the vortices.
We associate this smoothness of the bulk fields away from the vortex
cores with the absence of short-wavelength sound waves. In fact, the
almost complete absence of such short-wavelength sound waves appears
to be characteristic for all quasi-stationary configurations that we
have observed in our simulations. (It also holds for the
quasi-stationary lattice configurations that can be constructed as
mentioned above.) In the evolution of our systems sound waves of this
kind are typically created in annihilation processes of vortices and
are then rather rapidly dissipated.\footnote{This can be easily seen
  in our movies of example evolutions, see footnote \ref{fn:movies}.}
Sound waves of short wavelength have a relatively high energy density.
According to the interpretation of the holographic coordinate $z$ as
an inverse energy variable \cite{Maldacena:1997re} one would expect
these sound modes to correspond to bulk excitations (more or less)
localised close to the boundary. Unfortunately, we did not succeed in
extracting this expected behaviour from our simulation data. As long
as the system still contains vortices, there are bulk excitations
spanning a wide range of wavelengths.  The said short-wavelength sound
waves are only a small contribution relative to this background, and
it is therefore difficult to isolate their contribution numerically.
Despite this numerical difficulty the bulk perspective offers an
attractive way of studying and interpreting the dynamics close to the
non-thermal fixed point.
%
%
\section{Summary and outlook}
\label{sec:summary}
In this work, we have studied the far-from-equilibrium dynamics of a
holographic superfluid at finite temperature and chemical potential.
In the holographic framework,
the classical solution of an Einstein--Maxwell--scalar
system in $(3+1)$ dimensions is dual to the quantum dynamics of a
$(2+1)$-dimensional superfluid. We have performed a numerical study
of the time evolution of the holographic superfluid in its superfluid phase
starting from a variety of far-from-equilibrium initial conditions corresponding
to quenches of the system. In particular, we have imprinted various
kinds of large ensembles of topological defects, \ie~quantised vortices
and anti-vortices, in the superfluid, and have evolved the resulting
non-equilibrium states for a long time. The evolution can be
interpreted in terms of an effective description in which vortices
and antivortices interact via sound waves. When vortex--anti-vortex
pairs annihilate energy is quickly dissipated into the heat bath.

At early times, the evolution of the system strongly depends on the initial
conditions, that is on the distribution and density of vortices and
anti-vortices.
We find that the system exhibits a new non-equilibrium universality
regime in the late-time stages of the evolution. Starting from any
of our initial conditions, that regime is entered at times $t\gtrsim 600$
in the units we have chosen. This universal regime is characterised by
a dilute `gas' of vortex defects. The system remains in the universal
regime until the final times $t_{\rm f} = 4000$ we observe in our
simulations. In order to obtain more insight into the dynamics
of the universal regime, we have analysed the time evolution of characteristic length
scales of the ensemble of topological defects, and have studied
spatial correlations of microscopic excitations via the occupation
number spectrum. We have observed scaling laws in these observables
in the universal regime of the system's evolution and have determined
the corresponding scaling exponents.
The power-law behaviour of the occupation number spectrum can be
related to turbulence. During the universal stage, the occupation
number spectrum scales as $n(k,t)\sim k^{-\zeta}$ in an
infrared-to-intermediate-momenta inertial range with scaling exponent
$4.1\lesssim\zeta\lesssim 4.3$.
This value is close to the scaling $n(k,t)\sim k^{-4}$
observed in classical statistical simulations of two-dimensional Bose
gases \cite{Nowak:2010tm,Nowak:2011sk} where it was shown to be
related to a dilute random distribution of vortices and
anti-vortices and, in turn, to so-called strong wave turbulence
\cite{Scheppach:2009wu}.

We have made the interesting observation that the evolution of the
system exhibits critical slowing-down during the universal regime.
This is a natural feature of non-equilibrium dynamics in the vicinity
of non-thermal fixed points. We have presented evidence supporting
this interpretation of the observed late-time universal dynamics of our superfluid.
We have hence found the first evidence for the presence of a
non-thermal fixed point in the dynamics of a holographic system.
This is particularly interesting as the occurrence of a non-thermal fixed point
indicates that the dynamics of the corresponding non-equilibrium state is universal and
independent of the microscopic details of the system. The same
universal dynamics is then expected to occur in a variety of other
quantum systems, thus connecting very different fields of physics, see for example \cite{Berges:2013fga,Berges:2014bba,Berges:2014yta}.
The observation of a non-thermal fixed point not only gives a new
view on the dynamics of the holographic superfluid.
The gauge/gravity duality also adds a new dimension to the understanding of
non-thermal fixed points, as it translates quantum dynamics in the
boundary theory to classical dynamics in the bulk. Thus, the duality
might provide a new avenue for an analytic treatment of non-thermal
fixed points, as these are dual to stationary solutions of nonlinear
partial differential but classical equations of motion.

At an intermediate stage (at about $200 \lesssim t \lesssim 400$)
during the evolution of our system we observe a power-law energy
spectrum which is consistent with Kolmogorov $5/3$-scaling.
Such behaviour has been reported before in \cite{Adams:2012pj} for
the same system that we study here.
However, we find Kolmogorov scaling only for 
part of the various initial conditions we have chosen. Specifically, 
Kolmogorov scaling emerges when the evolution starts from random 
distributions of vortex defects, while this is not the case when we 
choose vortex lattices as initial configurations, the type of initial conditions used in \cite{Adams:2012pj}.
At the intermediate times concerned here, the system is
still affected by the initial distribution of vortex defects, while soon
afterwards it enters the universal regime discussed above which
exhibits a different scaling behaviour. Kolmogorov scaling thus seems to occur
only in a transient way. However, it appears that details of the initial conditions
can influence the system for a sufficiently
long time to completely prohibit the emergence of a transient Kolmogorov scaling.
This may also depend on the way in which statistical noise is implemented
in the numerical simulation for particular sets of quenches.
Further study is needed to understand better the conditions for
the occurrence of Kolmogorov scaling.

There are many open questions to be addressed in order to 
obtain a full understanding of the holographic superfluid. 
It will certainly be very interesting to study how the properties and various 
observables of the system depend on temperature and chemical 
potential, both within and across the phase boundaries of the 
superfluid phase. Another interesting point for further investigation 
concerns the coupling strength of the holographic superfluid. In general, the holographic duality maps a weakly coupled  
classical gravity system to a strongly coupled quantum system. 
It would be useful to investigate what exactly that 
means in relation to other descriptions of superfluids 
which have the coupling as an explicit parameter. 
A detailed comparison of the behaviour of, for instance, 
typical length or time scales in the holographic superfluid and 
in semi-classical (Gross--Pitaevskii) descriptions of superfluidity 
can give insight into this problem. 
%
%
\begin{acknowledgments}
  We would like to thank Paul Chesler, Elias Kiritsis, Hayder Salman,
  and Hong-bao Zhang for useful discussions.
  C.\,E.\ and A.\,S.\ are grateful to the Mainz Institute for
  Theoretical Physics (MITP) for its hospitality and its partial
  support during the final stages of this work.
  This work was supported by the Alliance Program of the Helmholtz
  Association (HA216/EMMI), the Deut\-sche Forschungsgemeinschaft
  (GA677/7,8), and the Heidelberg Center for Quantum Dynamics.
  A.\,S.\ acknowledges support in the framework of the cooperation
  contract between the GSI Helmholtzzentrum f\"ur Schwerionenforschung
  and Heidelberg University.
  Our plots have been created with \texttt{matplotlib}
  \cite{Hunter:2007} and \texttt{Mayavi} \cite{Ramachandran:2011}.
\end{acknowledgments}
%
%
\appendix
\section{Equations of motion and their numerical implementation}
\label{app:NumImpl}

In this appendix, we present the explicit form of the equations of
motion of our $(3+1)$-dimensional holographic model and technical
details concerning their numerical solution. The model consists of an
Abelian Higgs model coupled to gravity on an AdS$_4$ spacetime with a
Schwarzschild black hole.

\subsection{Equations of motion and holographic dictionary}
\label{app:EOMs}
As discussed in \Sect{model}, we solve the equations of motion of the
matter part of the action \eq{47} which we quote here again for convenience:
\begin{equation}
  \label{eq:action-app}
  \begin{gathered}
    S = \frac{1}{2\kappa} \int \D^4 x \sqrt{-g}\left(\mathcal{R} + \frac{6}{\LAdS^2} + \frac{1}{q^2}\mathcal{L}_{\text{matter}}\right) \,,\\
    \mathcal{L}_{\text{matter}}= -\frac{1}{4} F_{MN}F^{MN} - \left\lvert(\nabla_M-\i A_M)\Phi\right\rvert^2 - m^2 \lvert\Phi\rvert^2.
  \end{gathered}
\end{equation}
We consider a fixed AdS$_4$-Schwarzschild background as defined in \Eq{metric}. Our general setup follows \cite{Adams:2012pj}.

The dual field theory can be tuned to be in the symmetry-broken phase by adjusting the horizon temperature and the chemical potential, if the mass $m$ is suitably chosen.
Our choice $m^2 = -2/\LAdS^2$ is within the range of permissible values.
For convenience we set $\LAdS \equiv 1$.

In deriving the equations of motion we use the probe limit of large charge $q$, in which the Einstein equations decouple from the equations for the Abelian Higgs model.
$q$ then drops out of these equations.
Denoting by $\nabla_M$ the metric covariant derivative and by $D_M =
\nabla_M - \i A_M$ the combined metric and gauge covariant derivative,
the equations of motion for $A_M$ and $\Phi$ take the form
\begin{align}
  \nabla_M F^{MN} &= J^N \,,\label{eq:2}\\
  \left(-D^2 + m^2 \right)\Phi &= 0 \,,\label{eq:3}
\end{align}
with the current
\begin{align}
  J^N = \i \left(\Phi^* D^N \Phi - \Phi \left(D^N \Phi\right)^* \right)\,.
  \label{eq:11}
\end{align}
We fix the gauge freedom by choosing the axial gauge, $A_z = 0$.

Our aim is to describe the dynamical evolution of inhomogeneous solutions of the above equations, describing, \eg, vortex excitations of the dual superfluid.
For this we begin by deriving static solutions spatially homogeneous in $x$ and $y$.
Taking the fields $A_M$ and $\Phi$ independent of the coordinates $x, y$, and $t$, we obtain the explicit equations of motion, second order in derivatives with respect to the holographic coordinate $z$.
For the gauge field, they read
\begin{align}
  0 &= z^2 A_t'' + 2\Im(\Phi'\Phi^*) \,,\label{eq:9}\\
  0 &= z^2(h A_i''+h'A_i') - 2\lvert\Phi\rvert^2 A_i \,,\label{eq:10}\\
  0 &= 2A_t\lvert\Phi\rvert^2 - \i h\left(\Phi^*\Phi'-\Phi^*{}'\Phi\right) \,, \label{eq:7}
\end{align}
where $i=x,y$ and the prime denotes a derivative with respect to $z$.
The last equation originates from the dynamic equation for $A_{z}$ and remains as a gauge constraint to ensure the chosen axial gauge.
For the scalar field we find
\begin{equation}
  \label{eq:12}
  0 = z^2 h \Phi'' - z \left(-2\i z A_t + 2h -z h'\right)\Phi'
      - \left(2\i z A_t - \i z^2 A_t' + z^2 \vec{A}^2 +m^2 \right)\Phi
\end{equation}
with $\vec{A} = (A_x, A_y)$.

The limiting value of the electrostatic component $A_t$ of the gauge field at
the conformal boundary $z=0$ sets the chemical potential in the dual
gauge theory,
\begin{equation}
  \label{eq:5}
  A_t(z) = \mu + \mathcal{O}(z) \,.
\end{equation}
Our units are fixed by setting $\Zh=1$ in the metric
\eqref{eq:metric}, which also fixes the black-hole temperature.  Then,
in these units the temperature in the boundary theory is $T =
3/(4\pi)$, and we further choose $\mu=6$, such that $\mu/T = 8\pi$.
With these choices the system is in the superfluid phase.  At the
black-hole horizon $z=\Zh$, we need $A_t(\Zh)=0$ for regularity of $A_M$. As we
do not want to switch on sources for the spatial parts of the $U(1)$
current dual%
\footnote{The expectation value $\langle j^\mu\rangle$ of the current
  dual to $A_M$ is defined via $\langle j^\mu\rangle = -\lim_{z\to
    0}\sqrt{-g}F^{z\mu}$, \cf~\cite{Adams:2012pj}.}
to $A_M$ because these would break isotropy, we impose vanishing of
$A_x, A_y$ at the boundary, and $A_x(\Zh) = 0 = A_y (\Zh)$ at the
horizon.
Close to the boundary, the scalar field $\Phi$ behaves as
\begin{equation}
  \label{eq:fieldoperatormap}
  \Phi(t, \vec{x}, z) = \eta(t,\vec{x}) z + \mathcal{O}(z^2) \,,
\end{equation}
with the source field $\eta$. Since we want the scalar operator $\psi$
to form a condensate due to spontaneous symmetry breaking we choose
$\eta=0$. Then, the expectation value $\langle\psi(t,\vec{x})\rangle$ of the operator dual to $\Phi$ can be identified with the coefficient of the quadratic term in the expansion,
\begin{equation}
  \label{eq:fieldoperatornosource}
    \Phi(t, \vec{x}, z) = \langle\psi(t,\vec{x})\rangle z^2 + \mathcal{O}(z^3) \,.
\end{equation}
To summarise, the boundary conditions for the gauge fields $A_\mu$
are
\begin{equation}
  \label{eq:45}
  \begin{aligned}
    A_t(z=0) &= \mu,\qquad A_t(z=\Zh) = 0 \,,\\
    A_x(z=0) &= 0,  \qquad A_x(z=\Zh) = 0 \,,\\
    A_y(z=0) &= 0,  \qquad A_y(z=\Zh) = 0 \,,
  \end{aligned}
\end{equation}
where $\mu$ is the chemical potential. There is one explicit boundary
condition for the scalar field $\Phi$, namely $\eta=0$, which can be
expressed as
\begin{equation}
  \label{eq:46}
  \left.\del_z\Phi(z)\right\rvert_{z=0} = 0\,.
\end{equation}
The second boundary condition for $\Phi$ is a behavioural one: $\Phi$
be regular at the horizon. Physically, this represents the infalling
boundary condition \cite{Son:2002sd}. We use these boundary conditions
both for the construction of the equilibrium solution and for the full
equations of motion.

The solution of Eqs.~\eqref{eq:9}--\eqref{eq:12} yields the
holographic representation of the thermal equilibrium configuration of
the superfluid.
Using \eqref{eq:fieldoperatornosource} we extract the order-parameter field
$\langle\psi\rangle$ from the solution.
With our choice of the parameters $\mu$ and $T$, and of units, the
equilibrium value for the density of the superfluid order parameter is
$n = \lvert\langle\psi\rangle\rvert^2 \approx 41.7$.

Next, let us consider the full equations of motion. To construct the
initial conditions, we perturb the homogeneous solution by putting
vortices on top of it at an initial time $t=\ti < 0$. For a vortex of
winding number $w$, this is done by imprinting its winding structure
onto the bulk scalar field $\Phi$, locally $\Phi(\ti,\vec{x},z) \to
\Phi(\ti,\vec{x},z) \cdot \e^{\i w \phi_{\text{v}}}$ for every
$z$-slice, where $\phi_{\text{v}}$ is a polar angle in the $xy$-plane,
centred on the respective vortex.
The bulk configuration representing a single vortex carries the phase
winding at every $z$-slice \cite{Keranen:2009re}.  This follows
naturally from the fact that all $z$-slices contribute to the dual
field configuration, and is consistent with continuity of the phase of
the bulk scalar field.
At the vortex positions $\vec{x}_{\text{v}}$ the scalar field has to
vanish to remain well-defined after the phase winding is imprinted, so
we set $\Phi(\ti,\vec{x}_{\text{v}},z)=0$ along all $z$.
At all other grid points we leave the absolute value of $\Phi$ at the
equilibrium value. Starting the simulation, the system very quickly
builds up stationary density profiles around the vortex cores,
approximately within 5 or 10 units of time for quench classes
$\mathcal{A}$ and $\mathcal{B}$ (see \Tab{init-conds}),
respectively. Therefore, we choose $\ti = -5$ and $-10$ for quench
classes $\mathcal{A}$ and $\mathcal{B}$, respectively. To induce
variations in the decay pattern of the vortex lattices (class
$\mathcal{B}$), we additionally perturb the phase with noise $\e^{\i
  \Re(\zeta(\vec{x}))}$ at time $t=0$ (and only there) when the vortex cores are fully
formed. $\zeta(\vec{x})$ is obtained as the inverse discrete Fourier
transform of $\zeta(\vec{k})$.
This is in turn constructed by populating the Fourier modes in a disk
of radius $N/100$ about the origin in momentum space with
$\mathcal{O}(1)$ complex Gau{\ss}ian noise, where $N$ is the number of
grid points along each of the directions $x,y$. All Fourier modes
outside this disk are set to zero.

We thus excite the system to a non-equilibrium state, as discussed in
\Sect{initialCond}. Solving the full set of equations of motion, we
follow the subsequent evolution of the superfluid order parameter
$\langle\psi\rangle$. At every time step, we extract it from the bulk
field $\Phi$ using \eqref{eq:fieldoperatornosource}.
It is convenient for numerics to rescale the scalar field $\Phi$ by
$1/z$ and work with $\Phit\equiv \Phi/z$. Using the `lightcone
derivative'
\begin{equation}
  \label{eq:41}
  \nabla_+X = \del_tX - \frac{h(z)}{2}\del_zX\qquad\text{for}\qquad X = A_x, A_y, \Phit
\end{equation}
and with $\vec{\nabla} = (\partial_x,\partial_y)$ we eventually obtain
the following system of equations to solve:
\begin{align}
  \partial_z^2A_t &= \partial_z\vec{\nabla}\cdot\vec{A} - 2\Im (\Phit^{\ast}\partial_z\Phit) \,,%
  \label{eq:35}\\[.6em]
  \partial_z\nabla_+A_x &= \frac{1}{2}\left(\partial_y^2 A_x%
    +\partial_x(\partial_zA_t-\partial_yA_y)\right)%
  - \lvert\Phit\rvert^2 A_x + \Im (\Phit^{\ast}\partial_x\Phit) \,,
  \label{eq:36}\\[.6em]
  \partial_z\nabla_+A_y &= \frac{1}{2}\left(\partial_x^2A_y%
    + \partial_y(\partial_zA_t-\partial_xA_x)\right)%
  - \lvert\Phit\rvert^2 A_y + \Im (\Phit^{\ast}\partial_y\Phit) \,,
  \label{eq:37}\\[.6em]
  \partial_z\nabla_+\Phit &= \frac{1}{2}\vec{\nabla}^2\Phit %
  -\i\vec{A}\cdot\vec{\nabla}\Phit + \i A_t\partial_z\Phit %
  -\frac{\i}{2}\left(\vec{\nabla}\cdot\vec{A} - \partial_zA_t \right)%
  \Phit-\frac{1}{2}\left(z + \vec{A}^2\right)\Phit \,.
  \label{eq:38}
\end{align}
%

\subsection{Numerical methods}
\label{app:NumMeth}

For the $z$-parts of the equations of motion, we use a collocation
method with a basis of Chebyshev polynomials on a Gau\ss--Lobatto grid
with 32 points in the holographic direction (see
\eg~\cite{Boyd2000}). After setting the horizon radius $\Zh = 1$, we
switch to a new coordinate $\tilde{z}\in [-1,1]$ defined by
\begin{align}
  z &= \frac{1}{2}(\tilde{z}+1) \,. \label{eq:22}
\end{align}
With respect to $\tilde{z}$, we work entirely in real space,
implementing $\del_{\tilde{z}}$-differentiation via matrix
multiplication.

We treat the directions $x$ and $y$ as periodic. This enables us to
use discrete Fourier transforms to efficiently compute $\del_x$- and
$\del_y$-derivatives. For the $(x,y)$ grid we choose a grid constant of
$a=1/3.5$ in the aforementioned units. All data shown in this paper
were produced on a $504\times 504\times 32$ grid ($x, y, z$-directions).

To compute the equilibrium configuration of the system we have to
solve the boundary-value problem defined by
\eqref{eq:9}--\eqref{eq:12}. We treat the nonlinearity of the
equations by using the Newton--Kantorovich iteration procedure.
Technically, we solve the resulting system of linear equations via an
LU decomposition with full pivoting, see
\eg~\cite{epperson2014introduction}.

For the full set of equations of motion we solve the boundary value
problems in \eqref{eq:36}, \eqref{eq:37}, and \eqref{eq:38} for
$\nabla_+A_x$, $\nabla_+A_y$, and $\nabla_+\Phi$, undo the shifts
\eqref{eq:41} to get the time derivatives, and use a
fourth-order/fifth-order Runge--Kutta--Fehlberg algorithm with
adaptive timestep size to propagate the fields one timestep
forward. We allow the timestep size $\tau$ to vary in the range $0.001 \le
\tau \le 0.1$ in our aforementioned units. We use
\eqref{eq:35} to update $A_t$ in every timestep.

During the simulations, in order to investigate the spatial
characteristics of the vortex ensembles, we determine the positions of
all vortices and anti-vortices in the superfluid order parameter
$\langle\psi(\vec{x},t)\rangle$ at every unit timestep. To this end,
we iterate over the whole $(x,y)$ grid, measuring the integrated phase
of the superfluid order parameter around each elementary plaquette.

\subsection{Performance}
\label{app:performance}

We implement the algorithm in C\verb!++!, using the \texttt{fftw3}
library \cite{Frigo05thedesign} for Fourier transformations and the
\texttt{Eigen} library \cite{eigenweb} for high-performance linear
algebra. We parallelise the numerical code for multicore architectures
with \texttt{OpenMP} \cite{Dagum:1998omp}.

Propagating a $352\times 352$ grid with 32 points in the holographic
direction up to time 600 utilising 4 threads on a regular desktop
computer with an Intel i7 processor takes approximately 8 hours.

Running on the Intel Xeon server architecture using 16 threads, a run
up to time 4000 on a $504\times 504$ grid with 32 points in the
holographic direction takes around 120 hours.

\section{Fitting of scaling exponents of occupation spectra}
\label{app:determ-scal-law}

Scaling laws of occupation number spectra take the simple form
$n(k)\sim k^{-\zeta}$ with scaling exponent $\zeta$.
In the case of the occupation number spectra discussed in
\Sect{critical-dynamics}, we fit the power law exponents. We employ
the Levenburg--Marquardt least-squares fitting algorithm to determine
these scaling exponents after choosing a momentum range for the fit by
eye.

There are two sources of uncertainty in this: the determination of the
momentum range and the uncertainty inherent in estimating the best fit
parameters. The fitting algorithm reports an uncertainty of typically
$0.03$ for the scaling exponents. Fixing the momentum range introduces
a larger source of uncertainty. To address this, we vary the endpoints
of the momentum range and thus determine a range of estimators for the
scaling exponent.
As a result of that, we estimate the uncertainty in our determination
of scaling exponents of the occupation number spectra to be
$0.1$. Therefore, we state results for fitted scaling exponents with
one decimal digit.

The data on the time dependence of our observables is noisier than the
data on occupation number spectra such that using a fitting algorithm
is not appropriate in this case. The uncertainty in our determination
of scaling exponents with respect to time is difficult to assess.


\providecommand{\href}[2]{#2}\begingroup\raggedright\endgroup

\end{document}